\documentclass[superscriptaddress,prx,amsmath,amssymb,nofootinbib,twocolumn,aps]{revtex4-2}
\usepackage[T1]{fontenc}
\usepackage{amsthm}
\usepackage{graphicx}
\usepackage[percent]{overpic}
\usepackage{times}
\usepackage{hyperref}
\hypersetup{
colorlinks=true,
linkcolor=red,
citecolor=red,}
\usepackage{textcomp}
\usepackage{enumerate}
\usepackage{multirow}
\usepackage{tabularx}
\usepackage[mathscr]{euscript}
\usepackage{mathtools}
\usepackage[version=4]{mhchem}
\usepackage[mathscr]{euscript}
\usepackage{dsfont}
\usepackage{amsfonts}
\usepackage[normalem]{ulem}
\usepackage{centernot}
\usepackage{stmaryrd}
\usepackage{longtable}
\usepackage{enumitem,kantlipsum}
\usepackage{array}
\usepackage{wrapfig}
\usepackage{tabu}
\usepackage{placeins}
\usepackage{enumitem}
\usepackage{dutchcal}
\usepackage{tabularx}
\usepackage[export]{adjustbox}
\usepackage{color,soul}
\usepackage{longtable}
\usepackage[dvipsnames]{xcolor}
\usepackage{nicematrix}

\usepackage{tikz}
\usetikzlibrary{quantikz}

\usepackage[most]{tcolorbox} % For styling the conversation boxes

\newtheorem{definition}{Definition}
\newtheorem{prop}{Proposition}
\newtheorem{lemma}{Lemma}
\newtheorem{theorem}{Theorem}
\newtheorem*{theorem*}{Theorem}
\newtheorem{cor}{Corollary}
\newtheorem{remark}{Remark}
\newtheorem*{ass*}{Assumption}
\newtheorem{ass}{Assumption}

\newtcolorbox{exbox}{
    colback=black!5, % Light blue background
    colframe=white, % Dark blue frame
    breakable % Allows box to break across pages
}
\theoremstyle{definition}
\newtheorem{example}{Example}

\newcommand\s[1]{_{\rm #1}}

\newcommand{\ketbra}[1]{ | #1 \rangle\!\langle #1 |}

\newcommand{\Tr} {\operatorname{Tr}}

\newcommand{\spn} {\operatorname{span}}
\newcommand{\rank} {\operatorname{rank}}
\newcommand{\conv} {\operatorname{conv}}

\newcommand{\inprod}[2] {\langle #1 , #2 \rangle}

\newcommand{\one}{\mathbf{1}}

\newcommand{\R}{\mathbb{R}}

\newcommand{\Ident} {\mathds 1}
\newcommand{\M}{{\cal M}}

\newcommand{\PP}{{\cal P}}
\newcommand{\Q}{{\cal Q}}

\newcommand{\HH}{{\mathscr H}}
\newcommand{\EE}{{\cal E}}

\newcommand{\St}{{\cal S}}
\newcommand{\V}{{\cal V}}

\newcommand{\B}{{\cal B}}

\newcommand{\OO}{{\cal O}}

\newcommand{\MM}{\mathsf{M}}
\newcommand{\pp}{\mathsf{P}}
\newcommand{\e}{\mathsf{E}}
\newcommand{\oo}{\mathsf{O}}
\newcommand{\ES}{{\cal ES}}
\newcommand{\RF}{{\cal RF}}

\newcommand{\trs}{^\mathsf{T}}

\def\Q{{\mathbb{Q}}}
\def\R{{\mathbb{R}}}

\def\min{\mathord{\rm min}}

\def\spn{\mathord{\rm span}}

\begin{document}

\title{A Unified Linear Algebraic Framework for Physical Models and Generalized Contextuality}
\author{Farid Shahandeh}
\affiliation{Department of Computer Science, Royal Holloway, University of London, United Kingdom}
\author{Theodoros Yianni}
\affiliation{Department of Computer Science, Royal Holloway, University of London, United Kingdom}
\author{Mina Doosti}
\affiliation{School of Informatics, Quantum Software Lab, University of Edinburgh, United Kingdom}

\begin{abstract}
We develop a bottom-up, statistics-first framework in which the full probabilistic content of an operational theory is encoded in its matrix of conditional outcome probabilities of events (COPE).
Within this setting, five model classes (preGPTs, GPTs, quasiprobabilistic, ontological, and noncontextual ontological) are unified as constrained factorizations of the COPE matrix.
We identify equirank factorizations as the structural core of GPTs and noncontextual ontological models and establish their relation to tomographic completeness.
This yields a simple, model-agnostic criterion for noncontextuality: an operational theory admits a noncontextual ontological model if and only if its COPE matrix admits an equirank nonnegative matrix factorization (ENMF).
Failure of the equirank condition in all ontological models therefore establishes contextuality.
We operationalize rank separation via two complementary methods provided by the linear-algebraic framework.
First, we use ENMF to interpret noncontextual ontological models as nested polytopes.
This allows us to establish that the boxworld operational theory is ontologically contextual.
Second, we apply techniques from discrete mathematics to derive a lower bound on the ontological dimensionality of COPE matrices exhibiting sparsity patterns, and use this bound to establish a new proof that a discrete version of qubit theory exhibits ontological contextuality.
By reframing contextuality as a problem in matrix analysis, our work provides a unified structure for its systematic study and opens new avenues for exploring nonclassical resources. 
\end{abstract}

\maketitle

\section{Introduction}

Quantum contextuality in its various formulations~\cite{Kochen1967,spekkens2005,Cabello2008,Abramsky2011} refers to the fact that any attempt to describe all quantum statistics classically, i.e., using a hidden-variable model, would require keeping track of the contexts in which each outcome was obtained.
It is therefore recognized as one of the most profound departures of quantum theory from classical formalism, with implications for many forms of quantum advantages, such as in computation~\cite{Raussendorf2013,Bermejo-Vega2017,Frembs2018,Shahandeh2021arXiv,Schmid2022}, communication~\cite{Montina2014,Ambainis2016,Gupta2023}, and state discrimination~\cite{Schmid2018}.
However, pinpointing the role of contextuality remains a formidable challenge~\cite{Meyer1999,Kent1999,Simon2000,Simon2001,Kirchmair2009,Pusey2018,Mazurek2016,Mazurek2021}.

A common obstacle with the conventional formalism is its reliance on a ``top-down'' approach that presupposes a significant structure. 
More precisely, generalized contextuality proofs typically begin by \textit{explicitly} identifying the statistically indistinguishable experimental procedures, followed by two alternative lines of analysis.
In the first, one shows that the set of constraints imposed by these indistinguishabilities, known as operational equivalences, is inconsistent in any ontological model~\cite{spekkens2005}.
In the second line, the indistinguishabilities are translated into inequalities that bound the statistics achievable in an ontological model satisfying the identities~\cite{Schmid2018Witness}.
In both approaches, the statistical indistinguishabilities may be obtained either directly from the experimental data or from identities implied by a generalized probabilistic theory (GPT). 

A conceptually distinct method from the two approaches above for detecting contextuality assumes a GPT. Then it checks whether its states and effects can be linearly embedded into classical state and effect spaces~\cite{Shahandeh2021,Schmid2021}.
The linear embeddability can be checked using a linear program~\cite{Selby2024LP}.
The approach has since been broadened to encompass embeddings of arbitrary GPTs into one another~\cite{Mueller2023}.

Such reliance on explicit statistical indistinguishabilities and on an underlying GPT obscures the direct connection between \textit{observable}\footnote{We would like to draw attention to the distinction between “observable” and “observed” as used throughout this paper.
In van~Fraassen's words~\cite{vanFraassen}, the former ``are not exhausted by those actually observed, nor even by those observed at some time, whether past, present, or future.''} statistics and the underlying nonclassicality.
It may also perpetuate the mistaken impression that (non)contextuality is a property of the GPT rather than a property of the observable statistics.
For example, it is easy to devise two GPTs that yield identical statistics, yet only one admits a linear embedding into classical state and effect spaces~\cite{Selby2024LP}.

In this work, we present a ``bottom-up,'' statistics‑first framework for the study of operational theories that yields a single, operationally transparent contextuality criterion we call the \textit{rank‑separation}.
To achieve this, we treat operational theories as languages with three distinct structures: operational, causal, and probabilistic.
The operational structure is understood as the syntax, while the causal and probabilistic structures are interpreted as the physical probabilistic semantics of the language. 
We then restrict our analysis to the simple prepare-measure causal structure and introduce the probabilistic structure in matrix form, which we call the \textit{matrix of conditional outcome probabilities of events} (COPE).
We then show that several physical models, as models of operational theories, emerge as different factorizations of the COPE matrix.
More specifically, we study five types of models arising as COPE matrix factorizations: (i) preGPT, (ii) GPT, (iii) quasiprobabilistic, (iv) ontological, and (v) noncontextual ontological.
While each of these is traditionally understood as either a ``theory'' or a ``model'' with different statuses, we think of them as representing the possible worlds permitted by the semantics of the operational theory~\cite{vanFraassen,Shahandeh2021}.
Thus, in the picture we present, all are treated on equal footing. 

There are several benefits to our approach.
First, it unifies all models at a conceptual level, making it possible to draw a clear distinction between assumptions pertaining to the operational theory and those pertaining to the models.
For example, we argue that different formulations of the tomographic completeness assumption are model-specific.
We then elucidate their connection to the probabilistic structure of the operational theory.

Building on this, the formalism we put forward recasts the landscape of models into a unified, purely linear algebraic framework, wherein each model class is defined by a precise set of constraints on a common matrix factorization.
This framework thus provides a powerful new tool for posing and analyzing foundational questions with precision and rigor. 
For instance, we show how the maps between models are explicit linear-algebraic transformations, making transparent how assumptions tighten, relax, or reshape the geometry of admissible models.

In addition, our framework establishes clear connections between foundational problems and frontier research in disciplines such as computational geometry and graph theory.
This connection enables us to leverage complexity results from these fields to investigate the computational complexity of obtaining different models, with a particular emphasis on noncontextual ontological models~\cite{yianni2025}.

Finally, while our approach focuses on the probabilistic structure of operational theories rather than on any particular experimental scenario and its \textit{observed} data, it can be applied directly to the latter.
In this way, our framework also provides a systematic means of determining which types of models a given statistical data set admits.
In particular, the rank‑separation criterion determines whether, within the scope of a given experiment, a noncontextual ontological model is possible, similar to the approach proposed in Ref.~\cite{Gitton2022}.
In this way, the same data-driven viewpoint provides a practical means of detecting nonclassicality in diverse experimental settings.

The paper is organized as follows.
In Sec.~\ref{sec:OP_formalism} we set out a refined formalism for operational theories that is tailored to our perspective and underpins the analysis to follow.
There, we focus on the probabilistic structure, introduce the COPE matrix, and state its underlying assumptions.
We use Spekkens' toy theory~\cite{Spekkens_toy} as an illustrative example. 
We then move on to define models of the probabilistic structure in Sec.~\ref{sec:models}.
For each model, we provide its structural definition, show its construction, and its equivalence to a factorization of the COPE matrix.
Spekkens' toy theory will continue to serve as an example of each procedure for each model class.
Most importantly, we establish two results.
First, all forms of tomographic completeness are model features.
Second, a property of the matrix factorizations---the \textit{equirank} condition---forms the cornerstone of noncontextuality in a broad sense.
Consequently, for ontological models, violation of the equirank condition, termed \emph{rank separation}, is both necessary and sufficient for contextuality. 
As part of our analysis, we further show the equivalence between the equirank condition and tomographic completeness.
With this groundwork laid, we proceed to Sec.~\ref{sec:Rank_sep}, where rank‑separation is put into action.
Through examples, we present two contrasting methods for proving rank‑separation: a geometric approach applied to the boxworld operational theory, and an ontic‑space dimensionality‑counting approach applied to the discrete-qubit operational theory.
In Sec.~\ref{sec:Conv}, we develop a general argument on the structure of maps between models, establishing that \textit{linear set‑valued} functions form the most general class of convexity‑preserving transformations. 
This insight, in turn, allows us to show how GPTs can be mapped into \textit{contextual} ontological models in the absence of any noncontextual ontological model.
Finally, Sec.~\ref{sec:concl} contains the concluding remarks.

\section{Formalism} \label{sec:OP_formalism}

The fundamental concept in this paper is representation-free \textit{operational theories}.
An operational theory is a collection of sentences over an operational language.
To describe them, we mainly use the existing terminology and notions introduced in Refs.~\cite{spekkens2005,Harrigan2010-ao} in combination with some ideas from Refs.~\cite{Chiribella2010,Chiribella2016}.
Each operational theory has three structures: an \textit{operational} one, a \textit{causal} one (including sequential and parallel compositions), and a \textit{probabilistic} one. 
Throughout this paper, we consider only the simplest causal structure of prepare-measure experiments on individual systems.

\subsection{Operational structure}

The primitive elements of operational theories are lists of laboratory instructions.
For example, ``turn the laser on AND adjust its intensity as such'' is an instruction for preparing a photonic system.
Similarly, propositions such as ``turn the detector on AND align its aperture as such'' associated with measurements of the system are measurement instructions.
The operational structure is composed of the prescriptions (propositions) for preparing and measuring the system, denoted by the sets $\PP:=\{\pp_i\}$ and $\M:=\{\MM_j\}$, respectively.
Each measurement then results in an outcome $\oo^j_k$ from a set of possible outcomes.
We denote the set of all outcomes of all measurements by $\OO$.
The sentences in the theory can graphically be represented as a circuit,
\begin{equation}\label{eq:QCS}
\begin{quantikz}[row sep=0.4cm]
\prepD{\{\pp_i\}_i} &  \meterT{\{(\MM_j,\oo^j_k)\}_{jk}} 
\end{quantikz},
\end{equation}
The semantics of such sentences are probabilistic as described next.

\subsection{Probabilistic structure}

Consider the sentence ``The sky is cloudy, so it will rain today.''
According to Jaynes~\cite{JaynesBook}, such a sentence is a simple declarative statement of facts, and it is only reasonable to assign a probability to the sentence itself.
This probability quantifies the plausibility of rain today, given the prior information that the sky is cloudy.
Similarly, ``Preparing a photon as such and measuring it as such will result in a vertical polarization'' is a sentence in the operational theory to which a degree of plausibility, i.e., probability, can be assigned. 

The probabilistic structure of an operational theory is thus given by a probability measure $p$ that takes every possible sentence (prepare-measure circuit) to a real number in the unit interval.
The image of $p$ is thus the set of conditional outcome probabilities for each sentence in the theory.
We can also think of this measure as a matrix $C$ the entries of which are the probabilities $p(k|\pp_i,\MM_j)$ of each outcome $\oo^j_k$ happening given the system's preparation $\pp_i$ and measurement $\MM_j$, i.e., $C$ is a matrix of \textit{conditional outcome probabilities of events} (COPE)\footnote{A COPE is, in fact, a 3-tensor. Here, we have flattened its indices corresponding to measurement settings and outcomes.}\footnote{A similar construction is widely used in communication theory: In the simplest scenario, every bipartite communication task can be formulated as two parties, Alice and Bob, collaboratively computing a function $f(x,y)$. 
This function is then arranged in a matrix form, called the \textit{communication matrix}.
The rows and columns of this matrix index Alice and Bob's actions, respectively.
The communication matrix is then used to analyze the communication complexity of the task.}\footnote{A COPE matrix is different from existing concepts such as probability matrices~\cite{Mazurek2016}, data tables~\cite{Selby2024LP}, and behaviors~\cite{Tsirelson1993} in what it represents and its purpose.
For example, a probability matrix presupposes a GPT, a data table refers to the data obtained in an experiment, and a behavior signifies the input-output correlations in a given experiment.
The COPE matrix, by contrast, encompasses the entire probabilistic structure of an operational theory.},
\begin{gather}\label{eq:COPE}
C := 
\begin{pmatrix}
\begin{array}{c}
   \begin{matrix}
p(1|\pp_1,\MM_1) & \cdots & p(1|\pp_I,\MM_1) \\
        \vdots &   & \vdots  \\
        p(n_1|\pp_1,\MM_1) & \cdots & p(n_1|\pp_I,\MM_1)
   \end{matrix}\\ 
   \hline
    \vdots \\
   \hline
   \begin{matrix}
        p(1|\pp_1,\MM_J) & \cdots & p(1|\pp_I,\MM_J)\\
        \vdots & & \vdots \\
        p(n_J|\pp_1,\MM_J) & \cdots & p(n_J|\pp_I,\MM_J)
   \end{matrix} \\
\end{array}    
\end{pmatrix}.
\end{gather}
Here, each outcome is mapped to a row, and each preparation is mapped to a column of $C$.
We have assumed that there are a total of $J$ measurements, with $n_j$ outcomes each, and $I$ preparations, so that $C$ is an $(\sum_{k=1}^J n_k)\times I$ matrix.
Note that, in general, $I$, $J$, and $n_j$ can be infinite.
We refer to the submatrix of $C$ containing the conditional probabilities of the $j$th measurement as $C^j$, i.e.,
\begin{gather}\label{eq:COPE_j}
C^j := 
\begin{pmatrix}
p(1|\pp_1,\MM_j) & \cdots & p(1|\pp_I,\MM_j) \\
        \vdots &   & \vdots  \\
        p(n_j|\pp_1,\MM_j) & \cdots & p(n_j|\pp_I,\MM_j)
\end{pmatrix}.
\end{gather}
The probabilistic structure also dictates that the outcome probabilities for each measurement given a preparation sum to one, that is, $\sum_{i} C^j_{ik}=1$.
In the language of Chiribella \textit{et al.}~\cite{Chiribella2010,Chiribella2016}, the COPE matrix is the matrix of the probability function that assigns a probability to each \textit{event} from the trivial system to itself.
We represent by $\mathcal{C}$ and $\mathcal{R}$ the set of all columns and rows of a COPE matrix, respectively, such that $\oo^j_k\mapsto C^j_{k:}$ and $\pp_i\mapsto C_{:i}$.
Here $C^j_{k:}$ and $C_{:i}$ denote the $k$th row of $C^j$ and $i$th column of $C$, respectively.

In studying operational theories, it is common to take a ``top-down'' approach in which abstract notions such as \textit{systems}, \textit{tests}, \textit{events}, \textit{states}, and \textit{effects} within a background \textit{operational probabilistic theory} (OPT) or a \textit{true} generalized probabilistic theory (GPT) are primitives.
For example, when it is said that, ``Spekkens' toy theory~\cite{Spekkens2009} is preparation noncontextual'' this usually means that the ontological representations of the theory's \textit{states} depend only on the operational equivalence class they represent.
It is also common to consider these states as a \textit{fragment} of the qubit state space with extremal pure states $\{\ketbra{0}, \ketbra{1}, \ketbra{+},\ketbra{-},\ketbra{+i}, \ketbra{-}\}$.
 
We deviate from the prevailing top‑down tradition by adopting a ``bottom-up''
methodology in which the probabilistic structure---the semantics for the sentences of the operational theory---is taken as the primary object of analysis, without presupposing any underlying theoretical primitives.
Within this framework, different classes of physical models appear as different factorizations of the probabilistic structure---the COPE matrix---placing all models of operational theories, including ontological models, on equal footing up to their metaphysical interpretations.
This recasts deep foundational questions into concrete problems in linear algebra.
A further virtue of our approach is its statistics‑first perspective, in which fundamental properties are inferred directly from the observable probabilistic structure.
Following Ref.~\cite{Shahandeh2021}, we refer to factorizations of the COPE matrix as \textit{models of the operational theory}.
We completely characterize the structural and analytical relationships between these models in Sec.~\ref{sec:models}.

Interpreting probabilities as extended logic~\cite{JaynesBook}, we make the following foundational assumption about the COPE matrix and its relation to the operational theory.
\begin{ass}\label{ass:COPE_comp}
    The COPE matrix contains the complete probabilistic structure of the operational theory.
\end{ass}

\noindent The COPE matrix is thus the bedrock of our formalism---the fundamental, model-independent representation of empirical reality from which all further analysis proceeds.
As an example, we have described Spekkens' toy theory in the COPE matrix formalism in Example~\ref{ex:Spekkens_COPE} below.
Indeed, there are no guarantees that a novel preparation (measurement) will never be identified.
We assume that the operational theory (and thus its COPE matrix) will be extended to include such futuristic procedures once they are discovered.

\begin{exbox}
\begin{example}\label{ex:Spekkens_COPE} 
\textbf{Spekkens' toy theory: Probabilistic structure.}
Imagine a world which accommodates an \textit{elementary} system:
There are six propositions describing how to prepare the system and three propositions for measuring it, each yielding two outcomes.
We thus have $\PP=\{\pp_1,\pp_2,\pp_3,\pp_4,\pp_5,\pp_6\}$, $\M=\{\MM_1,\MM_2,\MM_3\}$, and $\OO=\{\oo_1,\oo_2,\oo_3,\oo_4,\oo_5,\oo_6\}$.
It is then assumed that a knowledge balance principle holds:
The amount of knowledge one possesses about the (ontic) state of the system at each instant of time equals the amount of knowledge one lacks.
This describes the elementary system of Spekkens' toy theory~\cite{Spekkens_toy}.
By the knowledge balance principle, the probabilistic structure of this operational theory is given by the following COPE matrix:
\begin{gather}\label{eq:STT_COPE}
C\s{S} := 
\begin{pmatrix}
1 & 0 & \frac{1}{2} & \frac{1}{2} & \frac{1}{2} & \frac{1}{2}\\
0 & 1 & \frac{1}{2} & \frac{1}{2} & \frac{1}{2} & \frac{1}{2}\\
\frac{1}{2} & \frac{1}{2} & 1 & 0 & \frac{1}{2} & \frac{1}{2} \\
\frac{1}{2} & \frac{1}{2} & 0 & 1 & \frac{1}{2} & \frac{1}{2} \\
\frac{1}{2} & \frac{1}{2} & \frac{1}{2} & \frac{1}{2} & 1 & 0 \\
\frac{1}{2} & \frac{1}{2} & \frac{1}{2} & \frac{1}{2} & 0 & 1
\end{pmatrix}.
\end{gather}
\end{example}
\end{exbox}

Using the probabilistic structure above, it is possible to define a notion of \textit{operational equivalence} over an operational theory~\cite{Spekkens2009,Chiribella2010}:
Two preparations $\pp_i$ and $\pp_j$ are said to be operationally equivalent, denoted as $\pp_i\cong\pp_j$, if and only if they induce the same probability distribution over all possible outcomes, i.e., $p(l|\pp_i,\MM_k)=p(l|\pp_j,\MM_k)$ for all outcomes of all measurements $\oo^k_l \in \OO$.
Similarly, two measurement outcomes $\oo^i_j$ and $\oo^k_l$ are said to be operationally equivalent, denoted as $\oo^i_j\cong\oo^k_l$, if and only if they occur with the same probability in all possible preparations, i.e., $p(j|\pp,\MM_i)=p(l|\pp,\MM_k)$ for all preparations $\pp \in \PP$.
Two measurements $\MM_i$ and $\MM_j$, on the other hand, are said to be operationally equivalent, denoted as $\MM_i\cong \MM_j$ if and only if there is a bijection $b:\MM_i\to \MM_j$ such that $b(\oo^i_k)\cong\oo^i_k$ for all $\oo^i_k$.
It follows that two operationally equivalent preparations give rise to two identical columns of the theory's COPE matrix, and two operationally equivalent outcomes result in two identical rows.
Two operationally equivalent measurements yield identical submatrices, $C^i=C^j$.

Assumption~\ref{ass:COPE_comp} implies that the converse of the above statements is also true, i.e., two identical columns of the COPE matrix denote two operationally equivalent preparations, two identical rows correspond to two operationally equivalent outcomes, and two identical measurement submatrices correspond to two operationally equivalent measurements.
This, in turn, leads to the following observation.

\begin{cor}\label{cor:IDRC}
    For two identical rows (columns) in the COPE matrix of an operational theory, there exist no preparations (measurement outcomes) that can separate them.
\end{cor}
\noindent Operational equivalence of outcomes, measurements, and preparations thus induces a trivial equivalence relation, the equality, on rows, measurement submatrices, and columns of the COPE matrix. 
As a result of Assumption~\ref{ass:COPE_comp} and Corollary~\ref{cor:IDRC}, operational equivalence classes of outcomes, $[\oo^j_k]$, measurements, $[\MM_j]$, and preparations, $[\pp_i]$ correspond to equivalence classes of rows, measurement submatrices, and columns of the COPE matrix under the equality relation, respectively. 
\begin{definition}\label{def:context}
    The elements of an operational equivalence class of recipes for outcomes, measurements, or preparations are called outcome, measurement, and preparation contexts, respectively.
    Equivalently, outcome, measurement, and preparation contexts can be considered to be the elements of an equality equivalence class of rows, measurement submatrix, and columns of a COPE matrix, respectively.
\end{definition}
\noindent The study of (non)contextuality is the study of the dependencies of models of operational theories on operationally equivalent recipes.
This is equivalent to analyzing how models of COPE matrices depend on the indices of identical rows and columns.

As illustrated in Example~\ref{ex:Spekkens_COPE}, we adopt an extremal‑quotiented convention for displayed COPE matrices.
A row (respectively a column) of $C$ is extremal if it cannot be written as a nontrivial convex combination of other rows (respectively columns) of $C$.
In the extremal‑quotiented convention, we display only extremal rows and columns.
The formal definitions and further discussion appear in Sec.~\ref{sec:IRC_rep}.

\begin{figure}
    \centering
    \includegraphics[width=0.8\linewidth]{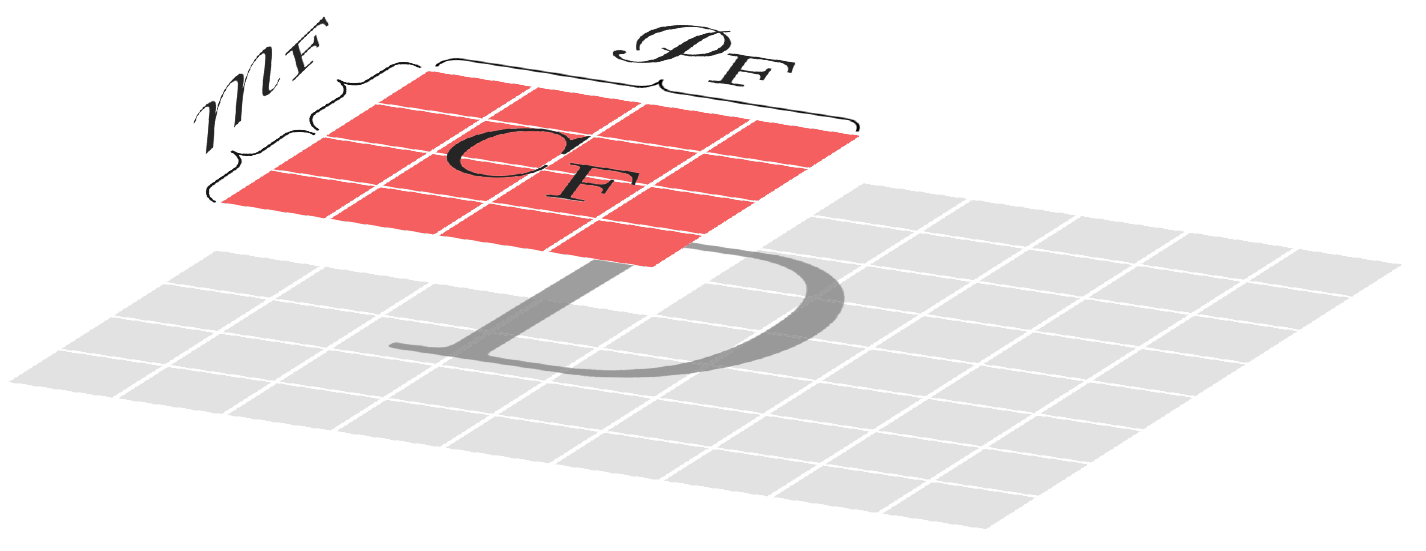}
    \caption{\textbf{Illustration of a fragment COPE matrix $C\s{F}$.} In considering fragments, the faithfulness assumption implies that the COPE matrix is a submatrix of the full operational theory's COPE matrix $D$ restricted to preparations $\PP_{\rm F}\subset\PP$ and measurements $\M_{\rm F}\subset\M$.
    Although the full COPE matrix $D$ satisfies Assumption~\ref{ass:COPE_comp}, its fragment $C\s{F}$ fails to do so.}
    \label{fig:faithful}
\end{figure}

It will also be useful for the following discussions and comparisons with other perspectives in the literature to introduce the concept of \textit{operational theory fragment}.
\begin{definition}\label{def:OT_fragment}
    An operational theory fragment (of a prepare-measure scenario) is given by a subset of all possible preparations of the operational theory, $\PP_{\rm F}\subset\PP$ and a subset of all possible measurements of the operational theory, $\M_{\rm F}\subset\M$.
    The probabilistic structure of the operational theory fragment is given by a probability data table $C_{\rm F}$ which is a submatrix of $C$.
    For consistency, we refer to this probability data table as the fragment COPE matrix.
\end{definition}
\noindent Note that an operational theory fragment is only meaningful with respect to a known (usually larger) operational theory.
In addition, a fragment COPE matrix clearly does not satisfy Assumption~\ref{ass:COPE_comp}.
Instead, the literature often implicitly makes an assumption, which we call faithfulness.
\begin{ass} \textbf{(Faithfulness).} \label{ass:faithful}
    If $C\s{F}$ is a fragment COPE matrix of an operational theory fragment, a container operational theory with a COPE matrix $D$ exists such that $C\s{F}:=D|_{\M\s{F},\PP\s{F}}$.
\end{ass}
\noindent We have illustrated this in Fig.~\ref{fig:faithful}.
Faithfulness is the implicit cornerstone of many contextuality analyses, e.g., whenever the sets of qubit density operators and POVMs are restricted to their subsets.
In the latter, the assumption is manifested in the use of quantum objects like density and nonnegative operators even though we only work with subsets of all states and all POVMs.
The reason for relying on this assumption is that, from a practical point of view, it is unreasonable to assume that one can operationally observe the entire COPE matrix.
Instead, the operational theory fragment is what one observes in a series of physical experiments or information-processing protocols.
Nevertheless, it is important to note that the models we construct in Section~\ref{sec:models} will not be founded on this assumption.
Therefore, we will not assume faithfulness in our examples either, unless explicitly stated.

\subsection{Representation of identical rows and columns of COPE matrices} \label{sec:IRC_rep}

There are two types of rows and columns in a COPE matrix of an operational theory: Extremal and convexly dependent.
In turn, identical rows and columns are either identical extremal or identical convexly dependent.
We also split operational recipes into two categories: Recipes whose corresponding rows, measurement submatrices, or columns are extremal are called extremal outcomes, measurements, or preparations, respectively.
The recipes that are not extremal are called convexly dependent.
We denote by ${\rm ex}\M\subset \M$ and ${\rm ex}\PP\subset \PP$ the sets of extremal measurements and preparations, respectively.
As noted in the previous section, we use an extremal convention, displaying only the extremal rows and columns of the COPE matrix.

Importantly, our extremal convention by itself does not yield an efficient representation.
To see this, consider preparing a photon in the horizontal polarization state $\ketbra{H}$ as follows.
Suppose we toss a coin and, depending on its outcome, either pass the photon through a polarizer, $\pp_1$, or reflect it off a polarizing beam splitter, $\pp_2$.
We denote the probabilistic recipe as $p\pp_1 + (1-p)\pp_2$\footnote{Here, the addition and multiplication must be understood as disjunction and conjunction of propositions, respectively~\cite{JaynesBook}.
In addition, similar to preparations, a probabilistic combination of two measurements $\MM_1$ and $\MM_2$ is denoted by $p\MM_1 + (1-p)\MM_2$.}.
The resulting statistics cannot be expressed as a nontrivial convex combination of other polarization statistics, yet the recipe itself is a probabilistic mixture of two distinct procedures.
In fact, there are infinitely many such recipes, one for each $p\in[0,1]$, all of which are statistically extremal. 
Including these in the COPE matrix would therefore be inefficient and, as we will show in the following sections, redundant.

We define $\cong_{{\rm ex}\PP}$ to be the restriction of the operational equivalence relation to extremal preparations ${\rm ex}\PP$, i.e., $\pp_1\cong_{{\rm ex}\PP} \pp_2$ if and only if $\pp_1$ and $\pp_2$ are extremal preparations and are operationally equivalent.
We call $\cong_{{\rm ex}\PP}$ the \textit{extremal-preparation equivalence relation}.
For measurements, we define $\cong_{{\rm ex}\M}$ to be the restriction of the operational equivalence relation to ${\rm ex}\M$ so that $\MM_1\cong_{{\rm ex}\M} \MM_2$ if and only if $\MM_1\cong\MM_2$ and they are extremal, i.e, their corresponding measurement submatrices in the COPE matrix are extremal.
We call $\cong_{{\rm ex}\M}$ the \textit{extremal-measurement equivalence relation}. 
Using Corollary~\ref{cor:IDRC}, extremal-preparation and -measurement equivalence relations induce equalities on the columns and rows of the COPE matrix.
\begin{definition}\label{def:COPE_quotient}
    The extremal quotiented COPE matrix $C\s{ex}$ is obtained by quotienting extremal rows and columns of $C$ with respect to $\cong_{{\rm ex}\PP}$ and $\cong_{{\rm ex}\M}$.
\end{definition}
\noindent In $C\s{ex}$, identical extremal columns and identical extremal measurements (with their respective sets of rows) appear only once.
In Sec.~\ref{sec:models}, we show that the representations of extremal recipes are unique across all models, including ontological models\footnote{In this sense, the (non)contextual properties of $C\s{ex}$ and $C$ are identical.}.
We therefore make the following assumption, unless stated otherwise.
\begin{ass}\label{ass:ext_quot}
    COPE matrices are extremal‑quotiented as in Definition~\ref{def:COPE_quotient}.
\end{ass}
\begin{definition}\label{def:ext_contexts}
    The elements of an equivalence class of preparation with respect to $\cong_{{\rm ex}\PP}$, or equivalently, the elements of an equality equivalence class of extremal columns of a COPE matrix, are called extremal preparation contexts.
    Extremal measurement contexts are defined as equivalent extremal measurement recipes with respect to $\cong_{{\rm ex}\M}$, or equivalently, identical sets of rows defining extremal measurements in the COPE matrix.
\end{definition}

\begin{figure}
    \centering
    \includegraphics[width=0.9\linewidth]{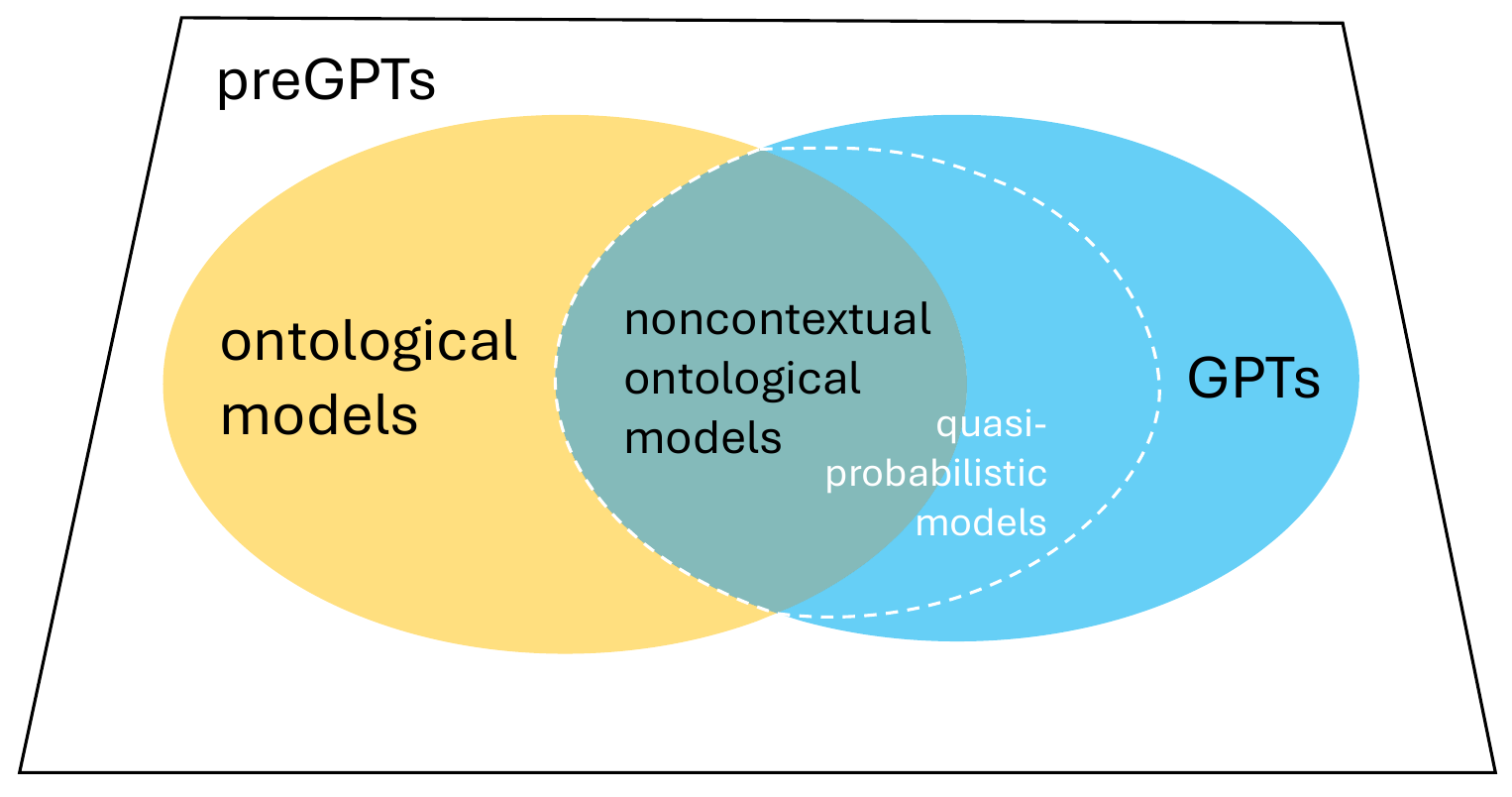}
    \caption{\textbf{Models of COPE matrices of operational theories.} Given the probabilistic structure of an operational theory encoded in its COPE matrix, one can universally obtain four types of linear models.
    Each model arises from a particular factorization subject to specific constraints:
    PreGPTs (no constraint), GPTs (quotiented), ontological models (nonnegative with $\one\trs$ as unit effect), and quasiprobabilistic models (quotiented with $\one\trs$ as unit effect).
    However, a COPE matrix may fail to admit a \textit{noncontextual} ontological model (quotiented \textit{and} nonnegative with $\one\trs$ as unit effect). 
    This contrasting class is the focus of our analysis in Secs.~\ref{sec:OM} and~\ref{sec:Rank_sep}.}
    \label{fig:Models}
\end{figure}

\section{Models of COPE matrices} \label{sec:models}

Recall that we assign a probability to the sentence ``The sky is cloudy, so it will rain today'' as its degree of plausibility~\cite{JaynesBook}.
Justifying this probability requires interpreting the sentence parts ``The sky is cloudy'' and ``it will rain today'' separately.
Similarly, in assigning a probability to ``Preparing a photon as such and measuring it as such will result in a vertical polarization'' as a sentence in the operational theory, ``Prepare a photon's polarization as such'' and ``measure the polarization of the photon as such'' require interpretations. 
In our framework for operational theories, such fine-grained interpretations, i.e. assigning mathematical objects to preparations, measurements, and their outcomes, belong to the realm of modeling, which is at a lower level compared to assigning probabilities to sentences.
In this section, we introduce five different types of models for the probabilistic structure of operational theories and analyze their relationships.

\subsection{Generalized probabilistic theories}

\subsubsection{Basic definitions and properties} \label{subsec:GPT_basics}

To build models, we require maps that send preparation and measurement outcome events to functions and define a probability rule using these functions.
Here, we introduce \textit{preGPTs}, which provide a general framework for achieving this.
The preGPT construction follows the familiar technical approach to GPTs~\cite{Shahandeh2021}.
It generalizes GPTs by relaxing the quotienting property in the sense that a preGPT provides room for assigning different mathematical objects (vectors) to statistically indistinguishable procedures.

A preGPT is the tuple $(\EE_{\rm pGPT},\St_{\rm pGPT},\V,\inprod{\cdot}{\cdot})$ where $\V$ is an ordered inner-product vector space, $\inprod{\cdot}{\cdot}$ is the inner product, and $\EE\s{pGPT}$ and $\St\s{pGPT}$ are constructed as follows.
Each possible event of the $j$th measurement, $\oo^j_k$, is represented by a vector $E(\oo^j_k)$ the collections of which for a fixed measurement form a {\it probability vector-valued measure} (PVVM) satisfying
\begin{equation} \label{eq:PVVM}
\begin{split}
    & E(\oo^j_k){\geqslant} 0 \quad \forall \oo_k\in\OO,\\
    & \exists! u \in \V \text{~s.t.~} \sum_{k} E(\oo^j_k){=}u,\\
    & E({\cup}_k \oo^j_k){=}\sum_k E(\oo^j_k).
\end{split}    
\end{equation}
Each vector $E(\oo^j_k)$ is called an effect, $u$ is called the unit element of the preGPT, and the collection of all operationally legitimate effects is denoted by $\EE_{\rm pGPT}$.
Further, by the convexity of the set of all events $\OO$ and the linearity of the map $E$, it is assumed that $\EE_{\rm pGPT}$ is convex. 
We denote the set of extremal points of $\EE_{\rm pGPT}$ by ${\rm ex}\EE_{\rm pGPT}$.

Now, let $\varrho$ be the function that maps preparations of the operational theory to linear functionals in $\V^*$, the dual of $\V$.
These vectors are called the preGPT states.
Using the Gleason's theorem~\cite{Shahandeh2021}, the image of $\varrho$ can be identified with a subset of the conic dual of $\EE_{\rm pGPT}$, 
\begin{equation} \label{eq:State_space}
    \St:=\{\rho\in\V{|}\inprod{E(\oo^j_k)}{\rho}\geqslant 0~\forall E(\oo^j_k)\in\EE_{\rm pGPT}{,}\inprod{u}{\rho}=1\}.
\end{equation}
Thus, denoting the state space of the preGPT by $\St_{\rm pGPT}$ and its extremal points by ${\rm ex}\St_{\rm pGPT}$, in general it holds that $\St_{\rm pGPT} \subseteq \St$. 
If $\varrho$ is surjective, i.e., for every vector in $\St$ there is a preparation procedure realizing it so that $\St_{\rm pGPT}=\St$, we say that the preGPT satisfies the no-restriction hypothesis.
Also, by the Gleason's theorem, the preGPT's probability rule is given by the inner product of $\V$, that is, for every entry of the COPE matrix, we have
\begin{equation} \label{eq:GPT_pob_rule}
    p(k|\pp_i,\MM_j) = \inprod{\varrho(\pp_i)}{E(\oo^j_k)}:=\inprod{\rho_i}{e^j_k}.
\end{equation}
We note that because the state space of a preGPT does not necessarily span $\V$, it is possible for multiple vectors in the image of $E$ to yield identical probabilities for all states.
As a result, there is enough room in the preGPT construction to map operationally equivalent outcomes and measurements to different effect vectors in the model.

It is also not difficult to imagine a construction, starting from the definition of states and the map $\varrho$, in which the resulting effects do not span $\V$ and thus, multiple vectors in the image of $\varrho$ may yield identical probabilities over all effects.
In this case, the preGPT construction allows for mapping operationally equivalent preparations to different state vectors within the model.

The ambiguities in the representations of equivalent operational procedures in a preGPT can be removed by imposing the following constraint.
\begin{ass} \label{ass:TC}
    The functions $\varrho$ and $E$ must be such that
    \begin{enumerate}[label=\roman*.]
        \item $\pp_i\cong\pp_j$ if and only if $\varrho(\pp_i)=\varrho(\pp_j)$, i.e., $\rho_i=\rho_j$.
        \item $\oo_i\cong \oo_j$ if and only if $E(\oo_i)=E(\oo_j)$, i.e., $e_i = e_j$.
    \end{enumerate}
\end{ass}
\noindent A preGPT which meets the above requirement is called a GPT.
Assumption~\ref{ass:TC} states that the GPT for an operational theory is a preGPT quotiented under the operational equivalence relation $\cong$ for outcomes and preparations.
That is, the states and effects of the GPT represent only equivalence classes of preparations and measurements~\cite{Chiribella2010}, or columns and rows of the COPE matrix.
From Corollary~\ref{cor:IDRC} and Assumption~\ref{ass:TC}, we conclude the following.
\begin{cor}\label{cor:ID_TC}
Any two rows (columns) in the COPE matrix of an operational theory which are identical are represented by the same effect (state) of the GPT.
\end{cor}

It is also worth highlighting that the notion of a preGPT is similar to that of a GPT fragment~\cite{Shahandeh2021,Selby2023fragments,Selby2024LP} in that they both reduce to GPTs when Assumption~\ref{ass:TC} is met.
Their difference, however, is that the domain of the definition of a preGPT is operational recipes, while a GPT fragment is a restriction of a background GPT.
Hence, while every GPT fragment is a preGPT (of some operational theory fragment---see Definition~\ref{def:OT_fragment}), the converse is not true.

PreGPTs also resemble \textit{operational probabilistic theories}~\cite{Chiribella2016}, which likewise allow for context-dependent representations of preparations and outcomes.
However, our preGPT construction shows that, in contrast to the common approaches, the linearity of such models does not follow from the quotienting operation alone.
We will clarify this in Sec.~\ref{sec:Conv}.

In Ref.~\cite{Shahandeh2021}, a notion of \textit{broad noncontextuality assumption} was introduced by extending Leibniz's principle of \emph{ontological} identity of \emph{empirical indiscernibles}~\cite{spekkens2019Leibniz} to all models of observable probabilities.
It states that \textit{any} model (ontological or otherwise) aiming to explain the probabilities must represent operationally equivalent preparations and outcomes in an identical and unique manner\footnote{One might argue that this is not precisely in accordance with Leibniz's principle; see Ref.~\cite{Adlam2021}.}.
In other words, the representation of preparations and outcomes in the model must only depend on their equivalence classes.
Thus, it can be said that two operational recipes result in identical probabilities because they are two realizations of the same causal mechanism\footnote{This is also a slight generalization of Spekkens' justification in Ref.~\cite{spekkens2005} for defining the notion of generalized noncontextuality of ontological models to all models.}.
It is then evident that Assumption~\ref{ass:TC} is the imposition of broad noncontextuality on preGPTs.

When constructing models of operational theories and their COPE matrices, we may require the model to possess specific properties.
An important example of such properties of models is \textit{tomographic completeness}.
We formulate it as follows.
\begin{definition} \textbf{(Tomographic Completeness of preGPTs).} \label{def:TC}
    A preGPT is tomographically complete if and only if:
    \begin{enumerate}[label=\roman*.]
        \item For any two states $\rho_i,\rho_j\in \St_{\rm pGPT}$, $\rho_i = \rho_j$ if and only if $\inprod{\rho_i}{e} =\inprod{\rho_j}{e}$ for all effects $e \in {\rm ex}\EE_{\rm pGPT}$.
        \item For any two effects $e_i,e_j\in \EE_{\rm pGPT}$, $e_i = e_j$ if and only if $\inprod{\rho}{e_i} = \inprod{\rho}{e_j}$ for all states $\rho \in {\rm ex}\St_{\rm pGPT}$.
    \end{enumerate}
\end{definition}
\noindent Note that, due to the convexity of the sets $\EE_{\rm pGPT}$ and $\St_{\rm pGPT}$, states and effects which induce identical distributions on boundaries of these sets induce identical distributions on the entire set.
We also have the following useful lemma.
\begin{lemma} \label{lemma:TC_LA}
    A preGPT is tomographically complete if and only if its sets of effects and states separate the points in the linear span of its states and effects, respectively.
\end{lemma}
\begin{proof}
    The if direction is immediate.
    We show the only if direction.
    By Definition~\ref{def:TC} (i), in a tomographically complete preGPT, effects separate states.
    We show that, as a result, the effects must also separate the entire ${\rm span}\{\St_{\rm pGPT}\}$.
    Suppose that this is not the case, i.e., there exists $g_1,g_2 \in {\rm span}\{\St_{\rm pGPT}\}$ such that $\inprod{g_1}{e}=\inprod{g_2}{e}$ for every $e\in {\rm span}\{{\rm ex}\EE_{\rm pGPT}\}$, but $g_1\neq g_2$.
    It follows that $\Delta g:= g_1-g_2 \in ({\rm span}\{{\rm ex}\EE_{\rm pGPT}\})^\perp$.
    Since $\Delta g \in {\rm span}\{\St_{\rm pGPT}\}$ there exists a linear combination of vectors such that $0\neq\sum_i \alpha_i \rho_i = \Delta g$ for $\rho_i\in{\rm span}\{\St_{\rm pGPT}\}$.
    It is therefore true that $\inprod{\sum_i \alpha_i \rho_i}{e}=\inprod{\Delta g}{e}=0$ for all $e \in {\rm span}\{{\rm ex}\EE_{\rm pGPT}\}$.
    Collecting all the terms in the linear combination with negative coefficients on one side, we have $\inprod{\sigma_1}{e} = \inprod{\sigma_2}{e}$ where $\sigma_1:=-(1/N)\sum_{i:\alpha_i<0} \alpha_i \rho_i$ and $\sigma_2:=(1/N)\sum_{i:\alpha_i>0} \alpha_i \rho_i$ where $1/N$ is a normalization factor.
    Note that, $\sigma_1$ and $\sigma_2$ are convex combinations of $\rho_i$s and thus $\sigma_1,\sigma_2\in\St_{\rm pGPT}$, i.e., they are valid states of the preGPT. 
    Moreover, $\sigma_1\neq\sigma_2$.
    This contradicts the assumption that ${\rm span}\{{\rm ex}\EE_{\rm pGPT}\}$ separates the points in ${\rm span}\{\St_{\rm pGPT}\}$, and in turn, the tomographic completeness.
    A similar argument can be applied to effects using condition (ii) of Definition~\ref{def:TC}.
\end{proof}

\noindent It is not difficult to see that the tomographic completeness mentioned above is equivalent to the quotienting assumption~\ref{ass:TC}.
\begin{prop} \label{prop:GPT_TC_preGPT}
    A GPT for an operational theory is a tomographically complete preGPT for that operational theory and vice versa.
\end{prop}
\begin{proof}
    We have,
    \begin{equation*}
    \begin{split}
        & \inprod{\rho_i}{e^k_l} = \inprod{\rho_j}{e^k_l} \quad \forall e^k_l \in {\rm ex}\EE_{\rm pGPT} \iff \\
        & \qquad \inprod{\varrho(\pp_i)}{E(\oo^k_l)}=\inprod{\varrho(\pp_j)}{E(\oo^k_l)}  \quad \forall \oo^k_l \in \OO \iff \\
        & \qquad  p(l|\pp_i,\MM_k) =  p(l|\pp_j,\MM_k) \quad \forall \oo^k_l\in\OO \iff \pp_i \cong \pp_j.
    \end{split}
    \end{equation*}
    where the third line is obtained using Eq.~\eqref{eq:GPT_pob_rule}.
    As a result, if (i) in Definition~\ref{def:TC} holds, then (i) in Assumption~\ref{ass:TC} holds and vice versa.
    A similar argument applies to effects and measurement outcomes.
\end{proof}

The implication of Proposition~\ref{prop:GPT_TC_preGPT} is that the assumption of tomographic completeness for preGPTs is no stronger than the fact that an operational theory provides a \textit{complete} description of the physical system (more precisely, a degree of freedom of a physical system) as stated in Assumption~\ref{ass:COPE_comp}.
As we will show shortly, every operational theory admits a tomographically complete preGPT, in other words, a GPT.
However, a stronger assumption of tomography can be found in the literature~\cite{Hardy2001,Barrett2007}.
We call this the \textit{fiducial tomographic completeness} to distinguish it from the weaker Assumption~\ref{ass:TC} (equivalently, Definition~\ref{def:TC}).
\begin{definition} \textbf{(Fiducial Tomographic Completeness).} \label{def:TC_MS}
    The strict subsets ${\rm ex}\EE_{\rm TOM}$ and $ex\St_{\rm TOM}$ of a GPT's extremal effects and states, ${\rm ex}\EE_{\rm GPT}$ and $ex\St_{\rm GPT}$, respectively, are fiduciary tomographically complete measurements and states if and only if:
    \begin{enumerate}[label=\roman*.]
        \item For any two states $\rho_i,\rho_j \in \St_{\rm GPT}$, $\rho_i = \rho_j$ if and only if $\inprod{\rho_i}{e} = \inprod{\rho_j}{e}$ for all effects $e \in {\rm ex}\EE_{\rm TOM}$.
        \item For any two effects $e_i,e_j \in \EE_{\rm GPT}$, $e_i = e_j$ if and only if $\inprod{\rho}{e_i} = \inprod{\rho}{e_j}$ for all states $\rho \in ex\St_{\rm TOM}$.
    \end{enumerate}
\end{definition}
\noindent The idea here is that, to know the exact state of the system in the model and the exact effect that occurred, one does not need to know the probability distribution induced over \textit{all} extremal measurement effects and the outcome probabilities for \textit{all} extremal states, respectively, as in Definition~\ref{def:TC}.
Such subsets, which make tomography possible, are called \textit{fiducial}~\cite{Hardy2001,Barrett2007}.
In this sense, fiducial tomographic completeness establishes the predictive power of a GPT by enabling the determination of the state (effect) vector from a subset of measurement outcomes (states) and the consistent prediction of probabilities in other experiments.
Note also that fiducial tomographic completeness is defined with respect to GPTs rather than preGPTs.
The reason is that, in a tomographically incomplete preGPT, even knowledge of the complete statistics across all states and effects is insufficient to identify them, let alone statistics from their subsets.

The uplifting of Definition~\ref{def:TC_MS} to COPE matrices is simple.

\begin{prop}
    An operational theory admits a fiduciary tomographically complete GPT if and only if 
    \begin{enumerate}[label=\roman*.]
        \item There exists a subset of outcomes ${\rm ex}\OO_{\rm TOM}\subset\OO$ such that for every two preparations $\pp_i$ and $\pp_j$ with $p(l|\pp_i,\MM_k) =  p(l|\pp_j,\MM_k)$ for all $\oo^k_l\in{\rm ex}\OO_{\rm TOM}$, we have $p(l|\pp_i,\MM_k) =  p(l|\pp_j,\MM_k)$ for all outcomes $\oo^k_l\in\OO$, that is, $\pp_i \cong \pp_j$.
        \item There exists a subset of preparations ${\rm ex}\PP_{\rm TOM}\subset\PP$ such that for every two outcomes $\oo^i_j$ and $\oo^k_l$ with $p(j|\pp,\MM_i) =  p(l|\pp,\MM_k)$ for all $\pp \in {\rm ex}\PP_{\rm TOM}$, we have $p(j|\pp,\MM_i) =  p(l|\pp,\MM_k)$ for all preparations $\pp\in\PP$, that is, $\oo^i_j\cong \oo^k_l$.
    \end{enumerate}
\end{prop}
\begin{proof}
    For the only if direction, suppose the operational theory admits a fiduciary tomographically complete GPT and let ${\rm ex}\OO_{\rm TOM}$ be the outcomes in the preimage of ${\rm ex}\EE_{\rm TOM}$.
    Then,
    \begin{equation*}
    \begin{split}
        & \rho_i = \rho_j \iff \inprod{\rho_i}{e^k_l} = \inprod{\rho_j}{e^k_l} \quad \forall e^k_l \in {\rm ex}\EE_{\rm TOM} \iff \\
        & \qquad \inprod{\varrho(\pp_i)}{E(\oo^k_l)}=\inprod{\varrho(\pp_j)}{E(\oo^k_l)}  \quad \forall \oo^k_l \in {\rm ex}\OO_{\rm TOM}
    \end{split}
    \end{equation*}
    Moreover, from the definition of GPTs (Assumption~\ref{ass:TC}),
    \begin{equation*}
        \rho_i = \rho_j \iff \pp_i \cong \pp_j.
    \end{equation*}
    Therefore,
    \begin{equation*}
        p(l|\pp_i,\MM_k)=p(l|\pp_j,\MM_k)  \quad \forall \oo^k_l \in {\rm ex}\OO_{\rm TOM} \iff \pp_i \cong \pp_j.
    \end{equation*}
    A similar argument applies to effects by defining ${\rm ex}\PP_{\rm TOM}$ to be the preimage of ${\rm ex}\St_{\rm TOM}$.
    
    The if direction is simply obtained by defining ${\rm ex}\EE_{\rm TOM}$ as the image of ${\rm ex}\OO_{\rm TOM}$ under $E$ and ${\rm ex}\St_{\rm TOM}$ as the image of ${\rm ex}\PP_{\rm TOM}$ under $\varrho$.
    Hence the result.
\end{proof}
\begin{cor}\label{cor:MSTC_rank}
    An operational theory with an extremal quotiented COPE matrix $C\s{ex}$ admits a fiduciary tomographically complete set of states if and only if the number of columns in $C\s{ex}$ is strictly greater than its rank.
    The latter is equivalent to $|{\rm ex}\St\s{GPT}|>\rank C\s{ex}$.
    Similarly, the operational theory admits a fiduciary tomographically complete set of effects if and only if the number of distinct rows in $C\s{ex}$ is strictly greater than its rank, or equivalently, $|{\rm ex}\EE\s{GPT}|>\rank C\s{ex}$.
\end{cor}
\begin{proof}
    This is immediate from Definition~\ref{def:TC_MS}.
\end{proof}
It is evident that Spekkens' toy theory is fiduciary tomographically complete, as its COPE matrix is $6\times 6$ of rank $4$.

So far, we have introduced preGPTs and GPTs as models for operational theories and studied two types of tomographic completeness for these models.
We have also provided the implications of tomographic completeness assumptions for operational theories and, in particular, their COPE matrices.
In the next section, we consider the problem of constructing preGPTs and GPTs given an operational theory.

\subsubsection{Construction of preGPTs and GPTs} \label{subsec:GPT_Const}

The fact that we can specify the probabilistic structure of operational theories as a matrix and both preGPTs and GPTs have linear structures allows us to provide a general strategy for constructing such models.
Our starting point is to notice the following.
\begin{prop}\label{prop:every_fac}
    Any factorization of a COPE matrix, $C=XY$, over reals in which the sums of rows of all $X^j$ submatrices corresponding to the $j$th measurement are identical and equal to a vector $u$ provides a valid preGPT.
\end{prop}
\begin{proof}
    This is immediate from the construction of preGPTs.
    Given $C=XY$, $\sum_i X^j_{i:} =u$ implies that every row of $X$ can be interpreted as a preGPT effect.
    Each column of $Y$ can then be interpreted as a preGPT state, as they correctly map effects to probabilities and their normalization is guaranteed by $1=\sum_i C^j_{ik} = \sum_i (X^jY)_{ik} = \inprod{u}{\rho}$.
\end{proof}

It is evident from Proposition~\ref{prop:every_fac} that the preGPT for an operational theory is not unique.
Thus, to demonstrate our methodology, we focus on the singular value decomposition (SVD), which is a well-known and efficient method for matrix factorization.
Beginning with the SVD of the COPE matrix into real-valued matrices $U \in \mathbb{R}^{m\times m}$, $\Sigma \in \mathbb{R}^{m\times n}$, and $V\in \mathbb{R}^{n\times n}$, such that $C=U\Sigma V$, we define $B:=\Sigma V$ so that $C=UB$ with $U$ being full-rank.
We also partition $U$ into $J$ submatrices $U^j$ such that $C^j=U^j B$.
Among all possible SVDs, we choose one such that the sums of rows of all $U^j$ submatrices are equal, i.e., $\sum_{i=1}^{n_j} U^j_{i:} = u$ for all $j$, where $U^j_{i:}$ is the $i$th row of $U^j$.
Note that the rows of $U$ span a vector space, call it $\V$, and the column space of $B$ is isomorphic to some subspace of $\V$, as required.
As a result, the rows of $U$ and columns of $B$ can be interpreted as effects and states of a preGPT, respectively, with $u$ being its unit vector.

Alternatively, we could define $A:=U\Sigma$ so that $C=AV$ and $V$ is full-rank.
This amounts to a preGPT in which the state space spans $\V$ and the span of effects is isomorphic to a subspace of $\V$.
To fully demonstrate this methodology, we continue with our Example~\ref{ex:Spekkens_COPE} and construct a preGPT for Spekkens' toy theory.

\begin{exbox}
\begin{example} \label{ex:Spekkens_preGPT}  
\textbf{Spekkens' toy theory: PreGPT.}
To obtain a preGPT for Spekkens' toy theory, we perform the SVD of its COPE matrix in Eq.~\eqref{eq:STT_COPE}, which leads to,
\begin{equation*}
\begin{split}
U\s{S} =
    \begin{pmatrix}
\frac{1}{\sqrt{6}} & 0 & 0 & -\frac{1}{\sqrt{2}} & 1 & -1 \\
\frac{1}{\sqrt{6}} & 0 & 0 & \frac{1}{\sqrt{2}} & -1 & 1 \\
\frac{1}{\sqrt{6}} & 0 & -\frac{1}{\sqrt{2}} & 0 & 1 & -1 \\
\frac{1}{\sqrt{6}} & 0 & \frac{1}{\sqrt{2}} & 0 & -1 & 1 \\
\frac{1}{\sqrt{6}} & -\frac{1}{\sqrt{2}} & 0 & 0 & 1 & -1 \\
\frac{1}{\sqrt{6}} & \frac{1}{\sqrt{2}} & 0 & 0 & -1 & 1
\end{pmatrix},
\end{split}
\end{equation*}
\begin{equation*}
\begin{split}
\Sigma\s{S} = 
\begin{pmatrix}
3 & 0 & 0 & 0 & 0 & 0 \\
0 & 1 & 0 & 0 & 0 & 0 \\
0 & 0 & 1 & 0 & 0 & 0 \\
0 & 0 & 0 & 1 & 0 & 0 \\
0 & 0 & 0 & 0 & 0 & 0 \\
0 & 0 & 0 & 0 & 0 & 0
\end{pmatrix},
\end{split}
\end{equation*}
\begin{equation}\label{eq:Spekkens_SVD}
\begin{split}
V\s{S}=
\begin{pmatrix}
\frac{1}{\sqrt{6}} & \frac{1}{\sqrt{6}} & \frac{1}{\sqrt{6}} & \frac{1}{\sqrt{6}} & \frac{1}{\sqrt{6}} & \frac{1}{\sqrt{6}} \\
0 & 0 & 0 & 0 & -\frac{1}{\sqrt{2}} & \frac{1}{\sqrt{2}} \\
0 & 0 & -\frac{1}{\sqrt{2}} & \frac{1}{\sqrt{2}} & 0 & 0 \\
-\frac{1}{\sqrt{2}} & \frac{1}{\sqrt{2}} & 0 & 0 & 0 & 0 \\
-\frac{1}{2} & -\frac{1}{2} & 0 & 0 & \frac{1}{2} & \frac{1}{2} \\
-\frac{1}{2\sqrt{3}} & -\frac{1}{2\sqrt{3}} & \frac{1}{\sqrt{3}} & \frac{1}{\sqrt{3}} & -\frac{1}{2\sqrt{3}} & -\frac{1}{2\sqrt{3}}
\end{pmatrix}.
\end{split}
\end{equation}
It thus follows that a preGPT can be given by $C\s{S}=U\s{S}B\s{S}$ where
\begin{equation}\label{eq:Spekkens_B}
\begin{split}
    B\s{S}=
    \begin{pmatrix}
\sqrt{\frac{3}{2}} & \sqrt{\frac{3}{2}} & \sqrt{\frac{3}{2}} & \sqrt{\frac{3}{2}} & \sqrt{\frac{3}{2}} & \sqrt{\frac{3}{2}} \\
0 & 0 & 0 & 0 & -\frac{1}{\sqrt{2}} & \frac{1}{\sqrt{2}} \\
0 & 0 & -\frac{1}{\sqrt{2}} & \frac{1}{\sqrt{2}} & 0 & 0 \\
-\frac{1}{\sqrt{2}} & \frac{1}{\sqrt{2}} & 0 & 0 & 0 & 0 \\
0 & 0 & 0 & 0 & 0 & 0 \\
0 & 0 & 0 & 0 & 0 & 0
\end{pmatrix}.
\end{split}
\end{equation}
The unit effect of this preGPT is given by the sum of the first, second, or third pair of rows of $U\s{S}$, which is $u\s{S}=(\sqrt{2/3},0,0,0,0,0)$.
The singular values show that the rank of $C\s{S}$ is four.
\end{example}
\end{exbox}

It is also easy to construct a GPT using the preGPT.
\begin{prop}\label{ref:pGPT_proj_GPT}
    Given a preGPT for a COPE matrix, $C=XY$, a projection onto the intersection of the row space of $X$ and the column space of $Y$ as $C=X\pi\pi'Y$ gives a GPT.
\end{prop}
\begin{proof}
    It is immediate that $\rank C =\rank X\pi =\rank \pi'Y$.
    Furthermore, we have that the row sums of $X'^j:=X^j\pi$ are identical across all measurements.
\end{proof}
\noindent In the construction using SVD of the COPE matrix above, a GPT can be obtained using the projections in Proposition~\ref{ref:pGPT_proj_GPT} as follows.
Suppose $r$ is the rank of $C$.
We define $M\in \mathbb{R}^{m\times r}$ by removing the $m-r$ right-most columns of $U$ and $S\in \mathbb{R}^{r\times n}$ by removing all-zero rows from $B$.
As above, partitioning $M$ into $J$ submatrices $M^j$ such that $C^j=M^j S$ and defining $M^j_{i:}$ as the $i$th row of $M^j$, we have that $\sum_{i=1}^{n_j} M^j_{i:} = u$ for all $j$ and some unit vector $u$.
Since the factorization $C=MS$ is a rank factorization, it is guaranteed that the rows of $M$ separate the columns of $S$ and vice versa.
This is precisely the GPT tomographic completeness condition of Lemma~\ref{lemma:TC_LA}. 
Hence, a GPT for an operational theory is given by
\begin{align}
        & C=MS \quad \textit{s.t.} \label{eq:C_mindec}\\
        & \rank C=\rank M=\rank S.\label{eq:GPT_equirank}
\end{align}
The construction above shows that the GPT for an operational theory is also not unique because the preGPT is not unique.
Finally, the dimensionality of the GPT is determined by the rank of the COPE matrix as stated in Eq.~\eqref{eq:GPT_equirank}, up to linear embeddings of $\V$ into larger vector spaces.
We call the relation between ranks of factors in Eq.~\eqref{eq:GPT_equirank} the \textit{equirank} condition.
\begin{cor}\label{cor:ID_GPT}
Every operational theory admits a GPT (tomographically complete preGPT) via the equirank real factorization of its COPE matrix $C=MS$ as in Eqs.~\eqref{eq:C_mindec} and~\eqref{eq:GPT_equirank}.
\end{cor}
\begin{cor}
    All GPTs for an operational theory are equivalent up to linear transformations.
\end{cor}
\begin{proof}
    We know from Corollary~\ref{cor:ID_GPT} that a GPT is an equirank real factorization of the COPE matrix of the operational theory as $C = MS$, where $M \in \mathbb{R}^{m\times k}$, $S \in \mathbb{R}^{k\times n}$, and $\rank(M) = \rank(S) = \rank(C) =: r$.
    Our claim is that any other GPT, given by $C = M'S'$ with $M' \in \mathbb{R}^{m\times k'}$ and $S' \in \mathbb{R}^{k'\times n}$ with $\rank(M') = \rank(S') = r$, can be obtained from the former by linear transformations $M' = M L_1$ and $S' = L_2 S$. 
    Indeed, from $C = MS=M'S'$ we have that every column of $C$ is a linear combination of the columns of $M$ or $M'$, so the column space of $C$ is a subspace of the column spaces of $M$ and $M'$.
    Now, since these spaces have equal dimensions, they must be identical.
    It is then immediate that there exists a map $L_1\in \mathbb{R}^{k\times k'}$ such that $M'=ML_1$.
    A similar argument applies to the row spaces of $C$, $S$, and $S'$ implying the existence of $L_2\in \mathbb{R}^{k'\times k}$ such that $S'=L_2S$.
\end{proof}
\begin{exbox}
\begin{example}\label{ex:Spekkens_GPT}
\textbf{Spekkens' toy theory: GPT.}
To transform the preGPT of Spekkens' toy theory in Example~\ref{ex:Spekkens_preGPT} into a GPT, we remove the unused subspace from the model.
To do so, we remove the last two columns of $U\s{S}$ in Eq.~\eqref{eq:Spekkens_SVD}, the entries of which are arbitrary and were chosen in such a way that the correct unit effect is obtained for all three measurements.
The state vectors can be obtained by removing the last two all-zero rows of $B\s{S}$ in Eq.~\eqref{eq:Spekkens_B}.
Thus,
\begin{equation}\label{eq:Spekkens_GPT}
\begin{split}
& M\s{S}=
\begin{pmatrix}
\frac{1}{\sqrt{6}} & 0 & 0 & -\frac{1}{\sqrt{2}}  \\
\frac{1}{\sqrt{6}} & 0 & 0 & \frac{1}{\sqrt{2}}  \\
\frac{1}{\sqrt{6}} & 0 & -\frac{1}{\sqrt{2}} & 0 \\
\frac{1}{\sqrt{6}} & 0 & \frac{1}{\sqrt{2}} & 0 \\
\frac{1}{\sqrt{6}} & -\frac{1}{\sqrt{2}} & 0 & 0 \\
\frac{1}{\sqrt{6}} & \frac{1}{\sqrt{2}} & 0 & 0 
\end{pmatrix},
\end{split}
\end{equation}
\begin{equation}
\begin{split}
S\s{S}=
\begin{pmatrix}
\sqrt{\frac{3}{2}} & \sqrt{\frac{3}{2}} & \sqrt{\frac{3}{2}} & \sqrt{\frac{3}{2}} & \sqrt{\frac{3}{2}} & \sqrt{\frac{3}{2}} \\
0 & 0 & 0 & 0 & -\frac{1}{\sqrt{2}} & \frac{1}{\sqrt{2}} \\
0 & 0 & -\frac{1}{\sqrt{2}} & \frac{1}{\sqrt{2}} & 0 & 0 \\
-\frac{1}{\sqrt{2}} & \frac{1}{\sqrt{2}} & 0 & 0 & 0 & 0 
\end{pmatrix}.
\end{split}
\end{equation}
The unit vector of this GPT is $u\s{S}=(\sqrt{2/3},0,0,0)$.
\end{example}
\end{exbox}

\subsection{Ontological models} \label{sec:OM}

\subsubsection{Basic definitions and properties}

An interesting class of models for operational theories is \textit{ontological models} in which an \textit{ontic} variable space $\Lambda$ is assumed to underlie physical phenomena.
Here, without loss of generality, we assume that there is a countable number of ontic points, hence $\Lambda$ is embedded in a vector space $\V\s{ont}\cong\R^s$ with $s:=|\Lambda|$.
We then need maps that send preparations and outcome events to their respective representations over $\R^s$.
This is done by sending ontic states of the system to the standard basis vectors in $\R^s$.
Then, since we may not know which \textit{true} ontic state was prepared in a preparation procedure $\pp$, these \textit{epistemic} states (ES) are represented by probability distributions over $\Lambda$, which are simply convex combinations of the standard basis vectors of $\R^s$. 
Let $\mu$ be the function that maps preparations to probability distributions over $\Lambda$.
Thus, we denote the epistemic state for preparation $\pp$ by $\mu(\pp)=(\mu_{\lambda}(\pp))_\lambda$ such that $0\leq\mu_{\lambda}(\pp)\leq 1$ and $\sum_\lambda \mu_{\lambda}(\pp)=1$ where the sum is over all ontic points.
It is assumed that the space of epistemic states is convex, i.e., if $\mu(\pp_i)$ and $\mu(\pp_j)$ are two epistemic states for any two preparations $\pp_i$ and $\pp_j$, the probabilistic combination of performing $\pp_i$ with probability $p$ and $\pp_j$ with probability $1-p$, represented by $p\pp_i + (1-p)\pp_j$, corresponds to $p\mu(\pp_i)+(1-p)\mu(\pp_j)$~\cite{spekkens2005}. 
We denote the closed convex set of all epistemic states by $\ES$, and its extremal points by ${\rm ex}\ES$. 

The outcome events in an ontological model are represented by nonnegative \textit{response} functions (RF) over $\Lambda$, i.e., by $\xi(\oo^j_k)=(\xi_{\lambda}(\oo^j_k)))_\lambda\trs$.
They satisfy $\xi_{\lambda}(\oo^j_k)\in[0,1]$ and $\sum_k\xi_{\lambda}(\oo^j_k)=1$ at any point $\lambda$ and for any measurement $\MM^j$.
Then, the probability of a particular outcome $k$ in the measurement $\MM_j$ given the preparation $\pp_i$ is obtained via total probability rule, 
\begin{equation} \label{eq:probrule_OM}
    p(k|\pp_i,\MM_j)=\sum_\lambda \xi_{\lambda}(\oo^j_k)\mu_{\lambda}(\pp) = \inprod{\mu(\pp_i)}{\xi(\oo^j_k)},
\end{equation}
where the inner product is that of $\R^s$.
Similar to states, the set of all response functions is a convex set denoted by $\RF$.
The subset of extremal response functions is ${\rm ex}\RF$.

It is possible to view ontological models as special cases of preGPTs where the vector space $\V$ is $\R^s$ and vectors for representing preparations and outcomes live in its positive orthant\footnote{Note that this only applies to finite-dimensional linear ontological models. 
The finite-dimensionality restriction can be lifted because every infinite-dimensional linear ontological model can be approximated to arbitrary precision with a finite-dimensional one.
However, there are nonlinear ontological models such as Bohmian quantum mechanics that do not fit into the above picture.}.
Then, the effects (response functions) become a subset of the standard unit hypercube, with its unit effect being the vector of ones $\one\trs$.
The normalized states (epistemic states) are a subset of the standard $(s-1)$-dimensional simplex whose extremal points are the standard basis vectors of $\R^s$.
Let us remark that, similar to a preGPT, an ontological model may assign different ontic representations to operationally equivalent procedures.
However, unlike preGPTs, it may not be possible to satisfy the no-restriction hypothesis for ontological models, hence their slightly different construction.
This is due to the nonnegativity constraint: If we start with a subset of the standard hypercube as the space of response functions, its dual as defined in Eq.~\eqref{eq:State_space} may not be contained in the positive orthant of $\R^s$.

In his seminal paper~\cite{spekkens2005}, motivated by Leibniz's principle of ontological identity of empirical indiscernibles~\cite{spekkens2019Leibniz}, Spekkens defined the phenomenon of contextuality as follows: ``A noncontextual ontological model of an operational theory is one wherein if two experimental procedures are operationally equivalent, then they have equivalent representations in the ontological model.''
Here, the justification is that since the ontological model provides a causal explanation for the data, it is reasonable to assume that two operational preparation recipes result in identical probabilities because they give rise to identical underlying epistemic states.
Similarly, two operational recipes for measuring the system yield outcomes with identical probabilities because they give rise to identical response functions.
\begin{ass} \label{ass:NC}
    The functions $\mu$ and $\xi$ must be such that
    \begin{enumerate}[label=\roman*.]
        \item $\pp_i\cong\pp_j$ if and only if $\mu(\pp_i)=\mu(\pp_j)$.
        \item $\oo_i\cong \oo_j$ if and only if $\xi(\oo_i)=\xi(\oo_j)$.
    \end{enumerate}
\end{ass}
\noindent An ontological model satisfying Assumption~\ref{ass:NC} is called a noncontextual ontological model\footnote{If only the equation for preparations is satisfied, the model is called \textit{preparation noncontextual}.
Similarly, if only the equation for outcome events is satisfied, the model is called \textit{measurement noncontextual}.}.
An operational theory that admits a noncontextual ontological model is called \textit{noncontextual}; otherwise, it is \textit{contextual}.

In parallel with preGPTs, we introduce a notion of tomography for ontological models as follows.
\begin{definition} \textbf{(Tomographic Completeness of Ontological Models).} \label{def:TC_OM}
    An ontological model is tomographically complete if and only if:
    \begin{enumerate}[label=\roman*.]
        \item For any two epistemic states $\mu_i,\mu_j\in \ES$, $\mu_i = \mu_j$ if and only if $\inprod{\mu_i}{\xi} =\inprod{\mu_j}{\xi}$ for all response functions $\xi \in {\rm ex}\RF$.
        \item For any two response functions $\xi_i,\xi_j\in \RF$, $e_i = e_j$ if and only if $\inprod{\mu}{\xi_i} = \inprod{\mu}{\xi_j}$ for all epistemic states $\mu \in {\rm ex}\ES$.
    \end{enumerate}
\end{definition}
\noindent A tomographically complete ontological model, similar to a tomographically complete preGPT, is a model such that its epistemic states and response functions are \textit{uniquely} identified via statistics.
An interesting observation that extends the similarities between ontological models and preGPTs in line with Proposition~\ref{prop:GPT_TC_preGPT}, hence further justifying our approach in the present paper, is as follows.
\begin{prop} \label{prop:OM_TC_NC}
    A noncontextual ontological model for an operational theory is a tomographically complete ontological model for that operational theory and vice versa.
\end{prop}
\begin{proof}
    We have,
    \begin{equation*}
    \begin{split}
        & \inprod{\mu_i}{\xi^k_l} = \inprod{\mu_j}{\xi^k_l} \quad \forall \xi^k_l \in {\rm ex}\RF \iff \\
        & \qquad \inprod{\mu(\pp_i)}{\xi(\oo^k_l)}=\inprod{\mu(\pp_j)}{\xi(\oo^k_l)}  \quad \forall \oo^k_l \in \OO \iff \\
        & \qquad p(l|\pp_i,\MM_k) =  p(l|\pp_j,\MM_k) \quad \forall \oo^k_l\in\OO \iff \pp_i \cong \pp_j,
    \end{split}
    \end{equation*}
    where the second line is obtained using Eq.~\eqref{eq:probrule_OM}.
    As a result, if (i) in Definition~\ref{def:TC} holds, then (i) in Assumption~\ref{ass:TC} holds and vice versa.
    A similar argument applies to effects and measurement outcomes.
\end{proof}
\noindent It is known that every operational theory admits an ontological model~\cite{Beltrametti1995}.
Thus, the foundational question is whether the model can be noncontextual.
Proposition~\ref{prop:OM_TC_NC} provides a deep interpretation of no-go theorems which rule out such models: The nonexistence of a noncontextual ontological model for an operational theory means that any ontological model for that theory is necessarily tomographically incomplete, i.e., it contains \textit{in-principle inaccessible} parameters. 

We further extend the similarities between preGPTs and ontological models by providing a statement analogous to Lemma~\ref{lemma:TC_LA}.
\begin{lemma} \label{lemma:TC_NC} 
    An ontological model is tomographically complete, and thus noncontextual, if and only if its sets of response functions and epistemic states separate the points in the linear span of its epistemic states and response functions, respectively.
\end{lemma}
\begin{proof}
The tomographic completeness of response functions, Definition~\ref{def:TC_OM} (i), means that, 
\begin{equation}\label{eq:contra_span_PNC}
    \mu_1 = \mu_2 \Leftrightarrow \forall f \in {\rm span}\{{\rm ex}\RF\} \quad \inprod{\mu_1}{f}=\inprod{\mu_2}{f}.
\end{equation}
We claim that Eq.~\eqref{eq:contra_span_PNC} holds if and only if the set of response functions separates the \textit{linear span of epistemic states}.
The ``only if'' direction is clear.
Thus, we prove the converse, namely that if Eq.~\eqref{eq:contra_span_PNC} holds, then the response functions separate the linear span of epistemic states.
Assume this is not true.
That is, there exist $g_1,g_2\in {\rm span}\{{\rm ex}\ES\}$ such that $g_1\neq g_2$ and $\inprod{g_1}{\xi}=\inprod{g_2}{\xi}$ for all $\xi\in {\rm ex}\RF$, and thus $\inprod{g_1}{f}=\inprod{g_2}{f}$ for all $f\in {\rm span}\{{\rm ex}\RF\}$.
It follows that $\Delta g:= g_1-g_2 \in ({\rm span}\{{\rm ex}\RF\})^\perp$.
Since $\Delta g \in {\rm span}\{{\rm ex}\ES\}$, it must be the case that some members of ${\rm span}\{{\rm ex}\ES\}$ have components in $({\rm span}\{{\rm ex}\RF\})^\perp$.
Since $\ES$ is bounded, closed, and convex, the latter implies that there exists a full-support distribution $\overline{\mu}$ in the interior of $\ES$ and a sufficiently small positive real number $\epsilon_0$ such that $\mu_\epsilon:=\overline{\mu} - \epsilon \Delta g$ is a valid epistemic state for all $0 <\epsilon < \epsilon_0$.
The members of the family of epistemic states $\{\mu_\epsilon\}_\epsilon$ satisfy $\inprod{\mu_{\epsilon_1}}{f}=\inprod{\mu_{\epsilon_2}}{f}=\inprod{\overline{\mu}}{f}$ for any $0<\epsilon_1<\epsilon_2 < \epsilon_0$ and for all $f\in {\rm span}\{{\rm ex}\RF\}$, while $\mu_{\epsilon_1}\neq \mu_{\epsilon_2}$.
This contradicts Eq.~\eqref{eq:contra_span_PNC} and the assumption that the model was tomographically complete.
A similar argument proves the claim for tomographic completeness of response functions, (ii), hence the result.
\end{proof}
In the next section, we show how one can construct ontological models using the probabilistic structure of the operational theory, i.e., its COPE matrix.

\subsubsection{Ontological models construction}

In Sec.~\ref{subsec:GPT_Const}, we showed that any SVD decomposition of the COPE matrix for an operational theory yields a preGPT.
We also showed that when the factorization is equirank, the preGPT will be tomographically complete, and thus we obtain a GPT.
We now show that a special factorization, known as the \textit{nonnegative matrix factorization} (NMF)~\cite{Wang2013,GillisBook}, applied to the COPE matrix of an operational theory yields an ontological model.
We then analyze the conditions for such ontological models to be noncontextual, hence providing a necessary and sufficient criterion for ontological noncontextuality of operational theories.
\begin{definition} \label{def:NMF}
    For any $m\times n$ (element-wise) nonnegative matrix $X$, its nonnegative matrix factorization (NMF) in $k$ dimension is a factorization of the form $X=WH$ wherein $W\in\R^{m\times k}_+$ and $H\in\R^{k\times n}_+$ are (element-wise) nonnegative matrices.
    It is also assumed that there are no all-zero rows and columns in $W$ and $H$, respectively.
    The smallest inner dimension in which such a factorization exists is called the nonnegative rank of $X$ and is denoted by $k_{\min}:=r_+$. 
\end{definition}
To construct an ontological model for an operational theory, let $C$ be its COPE matrix with $I$ preparations and $J$ measurements as in Eq.~\eqref{eq:COPE}.
Suppose that $C=WH$ is an NMF of $C$ in $k$ dimensions, where $k$ can be infinite.
Furthermore, suppose $W^1$ is the submatrix of $W$ corresponding to the submatrix $C^1$ of $C$ for the first measurement.
Defining $d_{j}:=\sum_i W^1_{ij}$, it is not difficult to verify that this sum always converges because $\sum_i C^1_{ij}$ for every column sum of $C^1$ converges to one.
In addition, $d_{j}>0$ as $W$ has no all-zero columns.
We now diagonally rescale the NMF as $C=WDD^{-1}H$, where $D:={\rm diag}[1/d_{j}]_{j}$ is a $k \times k$ diagonal matrix.
By defining $P:=D^{-1}H$, it is thus ensured that $P$ is column stochastic, i.e., its rows can be interpreted as probability vectors.

Next, we show that $R:=WD$ is a valid matrix of response functions.
Consider the fragment COPE matrix $C^j$, which corresponds to the $j$th measurement with $n^j$ (possibly countably infinite) outcomes.
Furthermore, let us denote the submatrix of $R$ corresponding to $C^j$ by $R^j$, that is $C^j=R^jP$.
We know that the sum of rows of $C^j$ is the row vector of ones, i.e., $\sum_{i}C^j_{i:}=\one\trs$.
Thus, using the fact that every column of $P$ is normalized to one, we find that $\sum_{i}R^j_{i:}=\one\trs$.
The latter implies that the columns of $R^j$ correctly represent a set of indicator functions for the $j$th measurement. 

Interestingly, it is not difficult to show that if we find an algorithm to construct ontological models for operational theories, we can use it to compute NMF of arbitrary nonnegative matrices.
To see this, suppose we want to compute an NMF for a nonnegative matrix $X$.
We can rewrite this as $X=XQ^{-1}Q:=YQ$ where $Q$ is a diagonal matrix having the sum of columns of $X$ as its diagonal entries so that $Y$ is nonnegative and column-stochastic.
We can now interpret $Y$ as the COPE matrix of a fictitious physical system and use our algorithm to compute an ontological model of it, $Y=RP$.
Finally, we get an NMF of $X$ using this ontological model factorization as $X=WH$ with $W:=R$ and $H:=PQ$.

It thus follows that the problems of finding an ontological model for an operational theory and NMF are equivalent.
Notably, in a similar approach, Harrigan \textit{et al.}~\cite{harrigan2008} introduced the concept of \textit{ontological factorization}, but they did not establish the equivalence between data factorization and NMF, nor its relation to generalized contextuality as discussed in this paper.

A particularly interesting question is: Given a COPE matrix, how can we compute its NMF and thus conceive an ontological model?
It is known that computing NMF for a given inner dimension $k$ is NP-hard~\cite{Vavasis2010,Shitov2017}.
The specification of the inner dimension is generally important, because every nonnegative matrix $C$ admits a trivial NMF, namely, $ C = C\Ident$, where $\Ident$ is the identity matrix.
In the language of ontological models, this amounts to saying that every COPE matrix admits a trivial ontological model, where the epistemic states are just the ontic states and the response functions are given by the rows of the COPE matrix~
\cite{Beltrametti1995,Janotta2014}.
As a result, finding \textit{nontrivial} ontological models for operational theories is a hard problem in general.

\begin{exbox}
\begin{example}\label{ex:Spekkens_COM}  
\textbf{Spekkens' toy theory: Contextual ontological model.}          
Spekkens' toy theory can be modeled by \textit{contextual} ontological models.
To give an example, consider the trivial model given by the trivial NMF of $C\s{S}$ of Eq.~\eqref{eq:STT_COPE}, i.e.,
\begin{equation}
\begin{split}
& W\s{S}=
\begin{pmatrix}
1 & 0 & \frac{1}{2} & \frac{1}{2} & \frac{1}{2} & \frac{1}{2}\\
0 & 1 & \frac{1}{2} & \frac{1}{2} & \frac{1}{2} & \frac{1}{2}\\
\frac{1}{2} & \frac{1}{2} & 1 & 0 & \frac{1}{2} & \frac{1}{2} \\
\frac{1}{2} & \frac{1}{2} & 0 & 1 & \frac{1}{2} & \frac{1}{2} \\
\frac{1}{2} & \frac{1}{2} & \frac{1}{2} & \frac{1}{2} & 1 & 0 \\
\frac{1}{2} & \frac{1}{2} & \frac{1}{2} & \frac{1}{2} & 0 & 1
\end{pmatrix},
H\s{S}=
\begin{pmatrix}
1 & 0 & 0 & 0 & 0 & 0 \\
0 & 1 & 0 & 0 & 0 & 0\\
0 & 0 & 1 & 0 & 0 & 0\\
0 & 0 & 0 & 1 & 0 & 0\\
0 & 0 & 0 & 0 & 1 & 0 \\
0 & 0 & 0 & 0 & 0 & 1
\end{pmatrix}.
\end{split}
\end{equation}
To verify that this model is indeed contextual, consider the following two preparations: (i) The balanced convex combination of the first and the second preparations, which in this model is represented by $\mu_{12}=(1/2,1/2,0,0,0,0)\trs$. (ii) The balanced convex combination of the third and the fourth preparations, which in this model is represented by $\mu_{34}=(0,0,1/2,1/2,0,0)\trs$.
These epistemic states induce the probabilities
\begin{equation}
    W\s{S}\mu_{12}=
    \begin{pmatrix}
1/2 \\
1/2\\
1/2\\
1/2\\
1/2\\
1/2
\end{pmatrix}=W\s{S}\mu_{34},
\end{equation}
showing that they cannot be distinguished by any of the possible measurements.
In other words, $\mu_{12}$ and $\mu_{34}$ are two distinct epistemic states in this model representing two operationally equivalent preparations.
Therefore, this model violates the noncontextuality Assumption~\ref{ass:NC}.
\end{example}
\end{exbox}

We now consider the class of noncontextual ontological models.
These can be easily characterized by applying Lemma~\ref{lemma:TC_NC} to the ontological model construction above.
\begin{theorem}\label{thm:NC_OM}
An operational theory admits a noncontextual ontological model if and only if its COPE matrix $C$ admits an equirank nonnegative matrix factorization (ENMF),
\begin{align}
        & C=RP \quad \text{such that} \label{eq:C_OM}\\
        & \rank C=\rank R=\rank P.\label{eq:OM_equirank}
\end{align}
\end{theorem}
\begin{proof}
    The fact that the NMF $C=RP$ is equivalent to an ontological model was shown before.
    The equirank condition of Eq.~\eqref{eq:OM_equirank} is immediate from Lemma~\ref{lemma:TC_NC}. 
\end{proof}

\begin{exbox}
\begin{example}\label{ex:Spekkens_NCOM}
\textbf{Spekkens' toy theory:
Noncontextual ontological model.}                                   
We now construct a noncontextual ontological model for Spekkens' toy theory.
It is important to note that, in sharp contrast to common approaches~\cite{Selby2024LP}, we do not start with a GPT looking for its embeddings into a simplex.
Instead, we start with the probabilistic structure of the operational theory, i.e., its COPE matrix given in Eq.~\eqref{eq:STT_COPE}, and consider its ENMF. 
There are no exact NMFs for $C\s{S}$ of Eq.~\eqref{eq:STT_COPE} in three dimensions because the nonnegative rank $r_+$ is always greater than or equal to the rank and, in this example, $r_+\geq 4$.
In four dimensions, however, we find,
\begin{equation}\label{eq:Spekkens_NCOM}
\begin{split}
R\s{S}=
\begin{pmatrix}
1 & 0 & 0 & 1 \\
0 & 1 & 1 & 0 \\
0 & 0 & 1 & 1 \\
1 & 1 & 0 & 0 \\
1 & 0 & 1 & 0 \\
0 & 1 & 0 & 1
\end{pmatrix}, 
\quad 
P\s{S}=
\begin{pmatrix}
\frac{1}{2} & 0 & 0 & \frac{1}{2} & \frac{1}{2} & 0 \\
0 & \frac{1}{2} & 0 & \frac{1}{2} & 0 & \frac{1}{2} \\
0 & \frac{1}{2} & \frac{1}{2} & 0 & \frac{1}{2} & 0 \\
\frac{1}{2} & 0 & \frac{1}{2} & 0 & 0 & \frac{1}{2}
\end{pmatrix},
\end{split}
\end{equation}
so that $C\s{S}=R\s{S}P\s{S}$.
We also see that $P\s{S}$ is column-stochastic.
It thus follows that this NMF provides an ontological model.
Furthermore, by $\rank{C\s{S}}=\rank R\s{S} = \rank P\s{S}=4$ the equirank condition of Eq.~\eqref{eq:OM_equirank} is satisfied and thus, the factorization given is an ENMF.
By Theorem~\ref{thm:NC_OM}, the model is noncontextual ontological.
Finally, Proposition~\ref{prop:OM_TC_NC} implies that the provided ontological model is also tomographically complete, that is, we can uniquely identify the model's epistemic states and response functions using the admissible measurements and preparations, respectively.
\begin{remark}
    We say Spekkens' toy theory is noncontextual because its COPE matrix, i.e., its probabilistic structure, admits a noncontextual ontological model as described above.
\end{remark} 
\end{example}
\end{exbox}

Equation~\eqref{eq:OM_equirank} relates the (non)contextuality property of all possible ontological models to a fundamental property of the operational theory, namely, the rank of its COPE.
The lesson we learn from Theorem~\ref{thm:NC_OM} is that, in contrast to the previous approaches, contextuality can be treated without an explicit analysis of operational equivalences among preparations and measurements, or operational identities between states and effects of a GPT.
The COPE formalism is thus ``GPT-agnostic'' in that it indicates whether a noncontextual ontological model of the probabilistic structure of the operational theory exists.
Such a generalization allows us to study contextuality in different tasks in a black box setting.
As we will show shortly, the above theorem puts forward a novel technique for proving contextuality by \textit{rank separation}, namely, 
by showing violations of Eq.~\eqref{eq:OM_equirank}.

From the above analysis, we also make the following useful observation, first established in Ref.~\cite{Selby2023}.
\begin{prop}\label{prop:COPE_normalization}
    Finding an ontological model (contextual or noncontextual) for a COPE matrix with $J$ measurements is equivalent to finding an ontological model for a COPE matrix with only one measurement.
\end{prop}
\begin{proof}
    Given a COPE matrix $C$ with $J$ measurements, we can normalize it as $C':=J^{-1}C$.
    An ontological model for $C'$ as $C'=R'P'$ gives an NMF for $C$ as $C=JR'P'$.
    This, in turn, gives an ontological model for $C$ as $C=RP$ where $R:=JR'$, and $P=P'$.
    
    Conversely, if we have an ontological model for $C$ as $C=RP$, it gives an NMF for $C'$ as $C'=J^{-1}RP$.
    This, in turn, gives an ontological model for $C'$ as $C'=R'P'$ where $R':=J^{-1}R$, and $P=P'$.
    
    Finally, $C'=R'P'$ satisfies the equirank condition of Eq.~\eqref{eq:OM_equirank} if and only if $C=RP$ does.
\end{proof}
We close this section with another useful corollary which is obtained by directly comparing Theorem~\ref{thm:NC_OM} with Eqs.~\eqref{eq:C_mindec} and~\eqref{eq:GPT_equirank}.
\begin{cor}\label{cor:NCOM_GPT}
    A noncontextual ontological model for an operational theory is equivalent to a GPT with nonnegative states and effects and the unit effect $u=\one\trs$.
\end{cor}
\noindent The difference between a GPT and an ontological model, in the form it is usually introduced and also presented here, is thus the nonnegativity of the representation and normality of the states, which allows one to interpret the model probabilistically, considering states as states of knowledge, and effects as response functions.

\subsection{Quasiprobabilistic models}

For completeness, we also mention a fifth class of models known as \textit{quasiprobabistic} models~\cite{Ferrie_2009,Schmid2020structure}.
We can think of them as models that lie between GPTs and ontological models in the following sense.
\begin{definition}\label{def:QP}
    A quasiprobabilistic model is a GPT (a tomographically complete preGPT) wherein the unit effect is given by vector of ones $\one\trs$.
\end{definition}
Having a unit vector of $\one\trs$ implies that for every state $q=(q_i)_i$ in a quasiprobabilistic model it holds that $\sum_i q_i =1$.
This ``normalization'' condition highlights their similarity with ontological models.
The radical difference between a quasiprobabilistic model and an ontological model, however, is that we trade nonnegativity of the latter with the tomographic completeness of the former.
In other words, quasiprobabilistic models are by definition tomographically complete and satisfy the equirank condition\footnote{Indeed, one can imagine quasiprobabilistic models which are not tomographically complete by considering preGPTs with unit vector $\one\trs$.
However, such models are not interesting because by relaxing the equirank condition as in preGPTs one can always construct ontological (probabilistic) models.}.
For this to be possible, the states and effects of the model are defined over all reals, i.e., they may attain negative values\footnote{In some literature, the phenomenon of negative distributions is accounted for by allowing for signed measures in probability theory~\cite{Blass_2021}.}.
This justifies the prefix \textit{quasi} in their name.

Finally, we have that any GPT can be transformed into a quasiprobabilistic model by solving the equation $\one\trs T^{-1}S=\one\trs$ for the invertible matrix $T$, where $S$ denotes the GPT states.
The quasiprobability effects and states then will be given by $M'=MT$ and $S'=T^{-1}S$.
The following proposition provides a simple method for finding a transformation $T$ that maps the GPT to a quasiprobabilistic model.
\begin{prop}\label{prop:GPT_quasi_T}
    Suppose that $C$ is a rank $r$ COPE matrix of some operational theory.
    Furthermore, let $C=MS$ be a GPT factorization of $C$ in $r$ dimensions.
    Let $S\s{TOM}$ be any submatrix of $S$ with $r$ linearly independent states.
    Then, $T=S\s{TOM}$ is a valid transformation from the GPT to a quasiprobabilistic model.
\end{prop}
\begin{proof}
Observe that $\one\trs T^{-1}S\s{TOM}=\one\trs \Ident=\one\trs$.
Hence, $T^{-1}=S\s{TOM}^{-1}$ correctly normalizes the subset of tomographic states.
Let us show that $T^{-1}$ also normalizes all the GPT states.
Consider an arbitrary state $s_i$ which can be written as a linear combination of tomographic states as $S\s{TOM}x=s_i$.
Suppose $M_j$ is the submatrix of $M$ for the $j$th measurement.
Because the outcome probabilities of $M_j$ sum to one, we have $\one\trs=\one\trs M_js_i=\one\trs M_jS\s{TOM}x=\one\trs x$.
The second equality follows because the rule of sum of probabilities also holds for the tomographic states, i.e., $\one\trs M_jS\s{TOM}=\one\trs$.
This means that the sum of the expansion coefficients of states in any GPT must obey $\one\trs x=\one\trs$.
Now, we have $\one\trs T^{-1}s_i=\one\trs S\s{TOM}^{-1}S\s{TOM}x=\one\trs x = \one\trs$, showing that every state in the GPT will be normalized by the above choice of $T$.
It is immediate that $M'=MT=MS\s{TOM}$ gives the correct matrix of effects for the quasiprobabilistic model.
\end{proof}
\noindent Note that by the above corollary, there are many solutions for $T$, meaning that there are many quasiprobabilistic models for a GPT.

\begin{exbox}
\begin{example} \label{ex:Spekkens_Quasi}   
\textbf{Spekkens' toy theory: Quasiprobabilistic model.}
We now demonstrate the possibility of a quasiprobabilistic model for Spekkens' toy theory.
While the noncontextual ontological model of Eq.~\eqref{eq:Spekkens_NCOM} is also a qusiprobabilistic model, our goal here is to show that there are quisiprobabilistic models that do not carry the ontological interpretation.
Thanks to Proposition~\ref{prop:GPT_quasi_T} and the GPT in Eq.~\eqref{eq:Spekkens_GPT}, this can be achieved using the invertible map,
\begin{equation}
    T:=
    \begin{pmatrix}
\sqrt{\frac{3}{2}} & \sqrt{\frac{3}{2}} & \sqrt{\frac{3}{2}} & \sqrt{\frac{3}{2}} \\
0 & 0 & -\frac{1}{\sqrt{2}} & \frac{1}{\sqrt{2}} \\
0 & -\frac{1}{\sqrt{2}} & 0 & 0 \\
-\frac{1}{\sqrt{2}} & 0 & 0 & 0
\end{pmatrix},
\end{equation}
to give the effect vectors $M'\s{S}:=M\s{S}T$ and state vectors $S'\s{S}:=T^{-1}S\s{S}$ as,
\begin{equation}\label{eq:Spekkens_quasi}
\begin{split}
    & M'\s{S}=
    \begin{pmatrix}
1 & \tfrac{1}{2} & \tfrac{1}{2} & \tfrac{1}{2} \\
0 & \tfrac{1}{2} & \tfrac{1}{2} & \tfrac{1}{2} \\
\tfrac{1}{2} & 1 & \tfrac{1}{2} & \tfrac{1}{2} \\
\tfrac{1}{2} & 0 & \tfrac{1}{2} & \tfrac{1}{2} \\
\tfrac{1}{2} & \tfrac{1}{2} & 1 & 0 \\
\tfrac{1}{2} & \tfrac{1}{2} & 0 & 1
\end{pmatrix},\\
& S'\s{S} = 
\begin{pmatrix}
1 & -1 & 0 & 0 & 0 & 0 \\
0 & 0 & 1 & -1 & 0 & 0 \\
0 & 1 & 0 & 1 & 1 & 0 \\
0 & 1 & 0 & 1 & 0 & 1
\end{pmatrix}.
\end{split}
\end{equation}
We see that, although the model of Eq.~\eqref{eq:Spekkens_quasi} shares its unit vector with the noncontextual ontological model of Eq.~\eqref{eq:Spekkens_NCOM}, it fails to be an ontological model.
This is because some of the state vectors in $S'\s{S}$ have negative entries, failing the respective columns to be probability distributions.
\end{example}
\end{exbox}

We close this section with the following observation.
\begin{lemma}\label{lemma:exC_models}
    The rows and columns of $C$ can be represented uniquely and identically in all five models.
\end{lemma}
\begin{proof}
    This is immediate from the constructions of the models discussed above.
    For any repeated extremal row or column, it suffices to reuse its corresponding vector representation in the model.
\end{proof}
\noindent It was in light of Lemma~\ref{lemma:exC_models} that we adopted Assumption~\ref{ass:ext_quot} in Sec.~\ref{sec:IRC_rep}, that the COPE matrices are extremal‑quotiented\footnote{Note that one can have preGPTs and ontological models that do not enjoy such a minimal quotienting and keep track of every possible context.}.
It also implies that the failure to represent \textit{convexly‑dependent} measurement and preparation contexts \textit{uniquely} and \textit{identically} in ontological models is precisely what necessitates contextuality.
We further demonstrate these arguments in an explicit example in Appendix~\ref{app:exBW}.

\section{Rank separation}\label{sec:Rank_sep}

In the previous sections, we discussed the operational formalism and its probabilistic structure encoded in the conditional outcome probabilities of events (COPE) matrix.
We also conducted a systematic comparative study of various models that can be devised to explain the statistics.
Our analysis drew a parallel between these models and their inherent assumptions and constraints.
In particular, we showed in Proposition~\ref{prop:GPT_TC_preGPT} that a tomographically complete preGPT is a GPT, and in Proposition~\ref{prop:OM_TC_NC} that a tomographically complete ontological model is a noncontextual ontological model.
We also showed that the assumption of tomographic completeness for both models is reflected as a simple constraint on their respective COPE matrix factorizations: The equirank condition in Eqs.~\eqref{eq:GPT_equirank} and~\eqref{eq:OM_equirank}.

While the equirank condition for a GPT can always be met, i.e., every operational theory admits a GPT, this is not generally true for ontological models.
Thus, the quest to prove that an operational theory does not admit a noncontextual ontological model reduces to establishing that the equirank condition of Eq.~\eqref{eq:OM_equirank} cannot be satisfied for any ontological model.
We refer to a failure of the equirank condition as \textit{rank separation}.

In the remainder of this paper, we employ two approaches to establishing rank separation.
First, by using the geometry of COPE matrices and noncontextual models as discussed in Sec.~\ref{subsec:Geometry}.
Second, through establishing lower bounds on the number of ontic states required in any ontological model of the operational theory and the minimum dimensionality of the vector subspace they span.
We demonstrate both these techniques using examples.

\subsection{Geometric approach: Boxworld}\label{sec:boxworld}

In our second example, we consider the boxworld, a hypothetical operational theory with four preparations and two dichotomic measurements.
The probabilistic structure of this theory is given by the COPE matrix,
\begin{equation} \label{eq:C_BW}
    C\s{BW}=
\begin{pmatrix}
1 & 0 & 0 & 1 \\
0 & 1 & 1 & 0 \\
1 & 0 & 1 & 0 \\
0 & 1 & 0 & 1 
\end{pmatrix}.
\end{equation}
As in the previous example, Spekkens' toy theory, let us begin with constructing a preGPT for this theory.
An SVD for $C\s{BW}$ is given by the matrices,
\begin{align}
    & U\s{BW}=
        \begin{pmatrix}
\frac{1}{2} & \frac{1}{2} & -\frac{1}{2} & -\frac{1}{2} \\
\frac{1}{2} & -\frac{1}{2} & \frac{1}{2} & \frac{1}{2} \\
\frac{1}{2} & -\frac{1}{2} & -\frac{1}{2} & -\frac{1}{2} \\
\frac{1}{2} & \frac{1}{2} & \frac{1}{2} & \frac{1}{2}
\end{pmatrix},\nonumber\\
    & \Sigma\s{BW} =
    \begin{pmatrix}
2 & 0 & 0 & 0 \\
0 & \sqrt{2} & 0 & 0 \\
0 & 0 & \sqrt{2} & 0 \\
0 & 0 & 0 & 0
\end{pmatrix},\displaybreak\nonumber\\
    & V\s{BW} = 
    \begin{pmatrix}
\frac{1}{2} & \frac{1}{2} & \frac{1}{2} & \frac{1}{2} \\
0 & 0 & -\frac{1}{\sqrt{2}} & \frac{1}{\sqrt{2}} \\
-\frac{1}{\sqrt{2}} & \frac{1}{\sqrt{2}} & 0 & 0 \\
-\frac{1}{2} & -\frac{1}{2} & \frac{1}{2} & \frac{1}{2}
\end{pmatrix}.
\end{align}
Note again that the last column of $U\s{BW}$ is arbitrary and is chosen such that it gives the correct unit vector for both measurements, namely, $u\s{BW}=(1,0,0,0)$.
These give the preGPT states as $B\s{BW}=\Sigma\s{BW}V\s{BW}$,
\begin{equation}
B\s{BW} =
    \begin{pmatrix}
1 & 1 & 1 & 1 \\
0 & 0 & -1 & 1 \\
-1 & 1 & 0 & 0 \\
0 & 0 & 0 & 0
\end{pmatrix},
\end{equation}
and the effects are $U\s{BW}$ so that $C\s{BW}=U\s{BW}B\s{BW}$.
The rank of $C\s{BW}$ is three, and our preGPT is four-dimensional, meaning that we can transform it into a GPT in lower dimensions.
As before, by removing the redundant column and row from $U\s{BW}$ and $B\s{BW}$, we find that the effects and states of our GPT to be
\begin{equation} \label{eq:BW_GPT}
    M\s{BW}=
        \begin{pmatrix}
\frac{1}{2} & \frac{1}{2} & -\frac{1}{2} \\
\frac{1}{2} & -\frac{1}{2} & \frac{1}{2} \\
\frac{1}{2} & -\frac{1}{2} & -\frac{1}{2} \\
\frac{1}{2} & \frac{1}{2} & \frac{1}{2}
\end{pmatrix},
S\s{BW} =
    \begin{pmatrix}
1 & 1 & 1 & 1 \\
0 & 0 & -1 & 1 \\
-1 & 1 & 0 & 0 
\end{pmatrix}.
\end{equation}
Similar to Spekkens' toy theory, the boxworld theory is fiduciary tomographically complete according to Corollary~\ref{cor:MSTC_rank} because its $4\times 4$ COPE matrix has a rank of $3$, implying that only a subset of its preparations and outcomes are sufficient to uniquely identify the GPT states and effects.

A quasiprobabilistic model for the boxworld can be obtained from its GPT in Eq.~\eqref{eq:BW_GPT} via the transformation
\begin{equation}
    T:= \begin{pmatrix}
1 & 1 & 1 \\
0 & 1 & 0 \\
0 & 0 & 1
\end{pmatrix}.
\end{equation}
This gives the effect vectors $M'\s{BW}:=M\s{BW}T$ and state vectors $S'\s{BW}:=T^{-1}S\s{BW}$ as,
\begin{equation}\label{eq:BW_quasi}
    M'\s{BW}=
    \begin{pmatrix}
\frac{1}{2} & 1 & 0 \\
\frac{1}{2} & 0 & 1 \\
\frac{1}{2} & 0 & 0 \\
\frac{1}{2} & 1 & 1
\end{pmatrix},
\quad
S'\s{BW} = \begin{pmatrix}
2 & 0 & 2 & 0 \\
0 & 0 & -1 & 1 \\
-1 & 1 & 0 & 0
\end{pmatrix}.
\end{equation}

We now address whether this operational theory is noncontextual.
We begin with constructing an ontological model.
Since $r_+\geq \rank C\s{BW}=3$, we check if an NMF in dimension three exists.
In the following, we present an intuitive argument that this theory does not admit a noncontextual ontological model.
But first, we need to take a detour to describe a geometric interpretation of ontological models.

\subsubsection{Geometry of ontological models}\label{subsec:Geometry}

\begin{figure}
    \centering
    \includegraphics[width=0.6\linewidth]{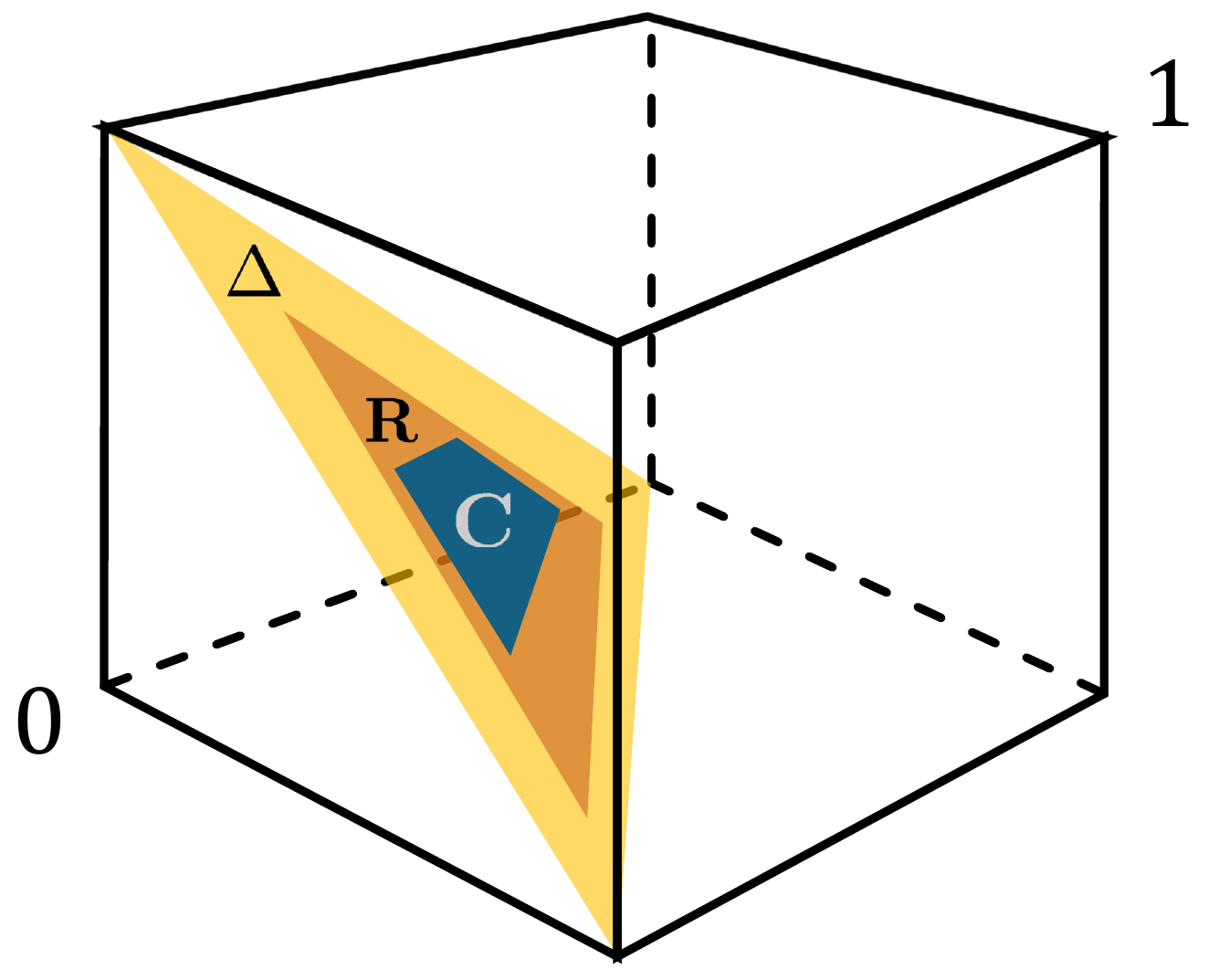}
    \caption{\textbf{Illustration of the geometry of noncontextual ontological models.} Factorizing the COPE matrix $C$ into nonnegative matrices of response functions, $R$, and epistemic states, $P$, gives an ontological model.
    Given the number of outcomes $Jn$, there is a simplex $\Delta\in\R^{Jn}$ which contains the polytope $\mathbf{R}$ constructed from the columns of $R$.
    The polytope $\mathbf{C}$ obtained from columns of the COPE matrix is obtained by convexly mixing the vertices of $\mathbf{R}$. The mixing weights are given by the epistemic states in $P$.
    The equirank condition of Eq.~\eqref{eq:OM_equirank} implies that in a noncontextual ontological model $\mathbf{C}$ and $\mathbf{R}$ are coplanar.}
    \label{fig:NCGeometry}
\end{figure}

Let us briefly sketch a geometric interpretation of ontological models.
This is schematically depicted in Fig.~\ref{fig:NCGeometry}.
Consider a COPE matrix $C$ for an operational theory.
Using Proposition~\ref{prop:COPE_normalization} we can assume that $C$ is column stochastic, i.e., it contains one measurement with $\sum_{j}n_j:=Jn$ outcomes.
As a result, we can assume each column of $C$ is a vector in the standard $(Jn-1)$-dimensional simplex whose extremal points are the standard basis vectors of $\R^{Jn}$.
Let us call this simplex $\Delta_{Jn}$ and the polytope corresponding to $C$ as $\mathbf{C}\subseteq\Delta_{Jn}$.
Now, consider an ontological model $C=RP$ for this operational theory in $s$ dimensions.
Since $C$ and $P$ are column-stochastic, $R$ is also column-stochastic.
We can thus interpret each of the $s$ columns of $R$ as a vector in $\Delta_{Jn}$ and represent it by a polytope $\mathbf{R}\subseteq\Delta_{Jn}$.
Each column of $P$ then provides the probability weights for mixing columns of $R$ to obtain the corresponding column of $C$. 
This implies that $\mathbf{C}$ is contained in $\mathbf{R}$.
Consequently, each ontological model corresponds to a sequence of nested polytopes, $\mathbf{C}\subseteq\mathbf{R}\subseteq\Delta_{Jn}$.

Note that, in the nested-polytopes picture above, the ontic dimension $s$ is reflected in the number of vertices required for the containment of $\mathbf{C}$. 
Hence, the minimal number of ontic states required for an ontological model as specified by the NNR of $C$ is equivalent to the minimum number of vertices defining a polytope $\mathbf{R}$ within $\Delta_{Jn}$ that contains $\mathbf{C}$.

What is the interpretation of a noncontextual ontological model from this geometric viewpoint?
For this, the equirank condition of Eq.~\eqref{eq:OM_equirank} together with the inclusion $\mathbf{C}\subseteq\mathbf{R}$ implies that the vertices of the polytopes $\mathbf{R}$ and $\mathbf{C}$ must span the same subspace.
That is, $\mathbf{R}$ cannot have vertices not within the span of the vertices of $\mathbf{C}$.  
The columns of $P$, however, do not admit a straightforward interpretation in terms of these polytopes. 
Instead, one can show, using an argument similar to the one above, that in a noncontextual ontological model, the rows of $P$ must span the same subspace as the rows of $C$.

\subsubsection{Proof that boxworld is ontologically contextual}

From above, we can interpret (the column-stochastic form of) $C\s{BW}$ as a polytope $\mathbf{C}\s{BW}$ in the standard 3-simplex $\Delta_4$ where every column of $C\s{BW}$ is a vertex of $\mathbf {C}\s{BW}$.
To find a noncontextual ontological model, our goal would be to find another polytope in $\Delta_4$ that lives in the same subspace as $\mathbf{C}\s{BW}$ and contains it.
However, looking at the zeros in each column of $C\s{BW}$, we notice that the vertices of $\mathbf{C}\s{BW}$ lie on the edges of $\Delta_4$ forming a square.
It is immediate that no intermediate polytope $\mathbf{R}$ with a larger or smaller number of vertices (i.e., a matrix of response functions in a smaller or larger ontic space) which lies in the subspace spanned by this square and $\mathbf{C}\s{BW}\subset\mathbf{R}\subseteq\Delta_4$ exists.
Therefore, there is a unique ontological model for the boxworld theory satisfying $\rank{W\s{BW}}=\rank{C\s{BW}}$.
This is given by the trivial model,
\begin{equation}\label{eq:BW_OM}
       W\s{BW}=
\begin{pmatrix}
1 & 0 & 0 & 1 \\
0 & 1 & 1 & 0 \\
1 & 0 & 1 & 0 \\
0 & 1 & 0 & 1 
\end{pmatrix},
\quad
H\s{BW} = 
\begin{pmatrix}
1 & 0 & 0 & 0 \\
0 & 1 & 0 & 0 \\
0 & 0 & 1 & 0 \\
0 & 0 & 0 & 1 
\end{pmatrix}.
\end{equation}
This model is clearly contextual because $\rank{H\s{BW}}>\rank{W\s{BW}}$.
As a result of its uniqueness, it follows that there is no ontological model for the boxworld that satisfies the equirank condition of Eq.~\eqref{eq:OM_equirank}.
Hence, the operational theory is ontologically contextual.

\subsection{Dimensional approach: Discrete qubit theory}

We now present a more operationally relevant example of how different types of models arise from a given operational theory.
We then use a dimensionality approach to prove its ontological contextuality.
We define the \textit{discrete-qubit operational theory} as follows.
We assume that there are two dense countable subsets of real numbers, $\R\s{pa}$ and $\R\s{ma}$, with index sets $\mathcal{I}_{\R\s{pa}}$ and $\mathcal{I}_{\R\s{ma}}$, respectively.
Lists of all admissible preparations is $\PP=\{\pp_i\}_{i\in\mathcal{I}_{\R\s{pa}}}$ and all dichotomic measurements is denoted by $\MM=\{\M_j\}_{\mathcal{I}_{\R\s{ma}}}$ with possible outcome events $\EE=\{\e^j_k\}_{(j,k)}$.
These preparations and measurements yield the probabilistic structure of the theory as an infinite COPE matrix $C\s{dq}$ which is a dense submatrix of the continuous COPE matrix of the qubit theory. 
Note, however, that we do not assume faithfulness~\ref{ass:faithful}: $C\s{dq}$ contains the complete probabilistic structure of the discrete-qubit operational theory (see Assumption~\ref{ass:COPE_comp}).
While the discrete-qubit operational theory is not identical to the qubit theory, the closures of its models are the standard qubit GPT, qubit quasiprobabilistic model, and qubit ontological model.

The reason for considering this approach is two-fold.
First, in any operational theory, it is only possible to realize a discrete set of preparations, measurement settings, and measurement outcomes.
The continuity of the \textit{states} and \textit{effects} is an idealization taking place at the level of abstract models.
For example, in classical mechanics, it is impossible to accommodate continuous, real-valued physical quantities operationally.
This, however, is not a problem because we think of the real-number classical mechanics as an abstract ontological model which is the closure of a discrete operational classical mechanics~\cite{Gisin_2019}.
Second, in an operational approach, we would not know in advance that the system is in fact an \textit{ideal} qubit.
All we can assume is that once enough data is collected, we will be able to model our preparations and measurements with a model, such as a GPT or an ontological model, as we will demonstrate shortly.
The closures of these models will then be the usual qubit GPT, quasiprobabilistic model, and ontological model.
This approach has been experimentally tested and its viability confirmed in Ref.~\cite{Mazurek2021}.

Following the discussion in Sec.~\ref{sec:IRC_rep}, we assume that the extremal rows and columns of this matrix (up to repetitions of equivalent outcomes in distinct measurements) are quotiented with respect to operational equivalence relations $\cong_{{\rm ex}\PP}$ and $\cong_{{\rm ex}\MM}$.
In an ideal world where preparations are extremal, and measurements are nonrefinable, $C\s{dq}$ has a rank of four~\cite{Mazurek2021} and can be factorized into
\begin{equation}
    M\s{dq}=
\begin{pmatrix}
    \begin{array}{c}
   \begin{matrix}
1 & e^1_{12} & e^1_{13} & e^1_{14} \\
1 & e^1_{22} & e^1_{23} & e^1_{24} \\
\hline
1 & e^2_{12} & e^2_{13} & e^2_{14} \\
1 & e^2_{22} & e^2_{23} & e^2_{24} 
   \end{matrix} \\
\hline
      \vdots
\end{array}   
\end{pmatrix},
\end{equation}
where $||\mathbf{e}^j_k||:=\sqrt{e_{k2}^{j2}+e_{k3}^{j2}+e_{k4}^{j2}}=1$ for all $j$ and $k$, and
\begin{equation}
    S\s{dq}=\frac{1}{2}
\begin{pmatrix}
    \begin{array}{cc}
   \begin{matrix}
1 & 1 & 1 & 1 \\
s^1_2 & s^2_2 & s^3_2 & s^4_2 \\
s^1_3 & s^2_3 & s^3_3 & s^4_3 \\
s^1_4 & s^2_4 & s^3_4 & s^4_4
   \end{matrix} 
      & \cdots \\
\end{array}   
\end{pmatrix},
\end{equation}
where $||\mathbf{s}^i||:=\sqrt{s_2^{i2}+s_3^{i2}+s_4^{i2}}=1$ for all $i$.
Here, the entries of $M\s{dq}$ and $S\s{dq}$ belong to the dense subsets $\R\s{oa}$ and $\R\s{pa}$, respectively, so that $C\s{dq}=M\s{dq}S\s{dq}$.
$M\s{dq}$ and $S\s{dq}$ are rank-four matrices and represent a GPT for the operational theory.
We emphasize that this GPT is not unique.

An equivalent formulation of the above GPT is to map its measurement effects and states via $\mathbf{e}^j_k\mapsto\Pi^j_k:=(\mathbf{e}^j_k\boldsymbol{\sigma}+\Ident)/2$ and $\mathbf{s}^i\mapsto\varrho_i:=(\mathbf{s}^i.\boldsymbol{\sigma}+\Ident)/2$, respectively, to operators on the Hilbert space $\HH_2$.
In this representation, the probability rule is given by the Hilbert-Schmidt product, $C^j_{ik}=\Tr(\Pi^j_k\varrho_i)$\footnote{Interestingly enough, this corresponds to another type of matrix factorization known as positive semidefinite (PSD) factorization~\cite{Fiorini2012,Vandaele2018}.}.
Finally, we note that given the above GPT and noting that the parameter sets $\R\s{oa}$ and $\R\s{pa}$ are dense subsets of $\R$, e.g., $\R\s{oa}=\R\s{pa}=\Q$, it is only natural to \textit{complete} the GPT by completing these subsets to the continuum $\R$, which results in the familiar qubit quantum mechanics.

A quasiprobabilistic model is now readily obtainable.
Suppose that we have recipes for preparing the qubit in four different ways corresponding to a rank-four submatrix of $S\s{dq}$ as\footnote{These correspond to the representation of, say, $\ketbra{0}$, $\ketbra{1}$, $\ketbra{+}$, and $\ketbra{+i}$.} 
\begin{equation}
    T = 
\begin{pmatrix}
1 & 1 & 1 & 1 \\
1 & -1 & 0 & 0 \\
0 & 0 & 1 & 0 \\
0 & 0 & 0 & 1 
\end{pmatrix}.
\end{equation}
We thus obtain a quasiprobabilistic model for the discrete-qubit operational theory by defining $T=S_4$ as in Proposition~\ref{prop:GPT_quasi_T} and
\begin{equation}
    M'\s{dq}:=M\s{dq}T, \quad S'\s{dq}:=T^{-1}S\s{dq}.
\end{equation}

Given $C\s{dq}$, an ontological model for the qubit can be given by $W\s{dq}:=C\s{dq}$ and $P\s{dq}:=\Ident$. 
In the continuous limit, the ontological space becomes identical to the space of positive operators on the Hilbert space $\Lambda=\B(\HH_2)$ so that the epistemic states will be given by $\mu(\lambda|\pp\s{ext})=\delta(\lambda-\psi_{\pp\s{ext}})$.
Here, $\pp\s{ext}$ is the operational equivalence class of the extremal preparation $\pp\s{ext}$ which corresponds to the pure quantum state $\ketbra{\psi_{\pp\s{ext}}}$.
It is also straightforward to derive the response functions for this choice of states.
The resulting ontological model is that of Beltrametti and Bugajski~\cite{Beltrametti1995}.

We now prove that the qubit theory does not admit a \textit{noncontextual} ontological model.
To do so, however, we need to take a detour proving lower bounds on the dimensionality of contextual and noncontextual ontological models.

\subsubsection{Dimensionality of ontological models}

\begin{theorem}\label{thm:Ontic_lowerbound}
    Suppose $C$ is a COPE matrix with $I$ preparations and $Jn$ measurement outcomes.
    Let $C_m$ be an $m\times m$ submatrix of $C$ such that each column (row) of $C_m$ has at least one zero entry that differs from all other columns (rows), uniquely identifying that column (row).
    Then,
\begin{enumerate}[label=(\alph{enumi})]
    \item The dimension of the ontic space $\V\s{ont}$ for any ontological model reproducing $C$ is lower bounded by the smallest integer $k$ such that $m\leq \binom{k}{\lfloor k/2\rfloor}$, and,
    \item The spans of the response functions and epistemic states satisfy $m\le
F\!\left(
\dim\spn\{\xi_i\},
\dim\spn\{\mu_i\}
\right)$, where $F(x,y):=\binom{x+y-4}{x-2}+2\binom{x+y-6}{x-3}$.
\end{enumerate}
\end{theorem}
\begin{proof}
(a) The vector space $\V\s{ont}\simeq\R^k$ underlies the ontological model.
Noting that the model is finite-dimensional, where every vector space is isomorphic to its dual, both epistemic states and response functions live in the same space $\V\s{ont}$.
We also assume that $m\geq 2$.
By assumption, for every two epistemic states $\mu_1$ and $\mu_2$ in the model, there is a response function $\xi_1$ such that $\inprod{\xi_1}{\mu_1}=0\neq\inprod{\xi_1}{\mu_2}$.
Similarly, there is a response function $\xi_2$ such that $\inprod{\xi_2}{\mu_1}\neq 0 =\inprod{\xi_2}{\mu_2}$.
Then, 1) it follows from nonnegativity of $\mu_i$ and $\xi_i$ ($i\in\{1,2\}$) that they live in two orthogonal subspaces, 2) the support of $\mu_i$ is not a subset of the support of $\mu_j$ for $i\neq j$, and, 3) the support of $\xi_i$ is not a subset of the support of $\xi_j$ for $i\neq j$.
Condition (2) (equivalently, (3)) implies that the ontic points must admit $m$ subsets such that none of them is a strict subset of another.

Now, there are $k$ ontic points to support the epistemic states and response functions.
Sperner's theorem states that for any set of size $k$, there can be a maximum of $K:=\binom{k}{\lfloor k/2\rfloor}$ subsets such that none of them contains another.
It is, therefore, clear that if $m>K$, it is impossible to fit the given epistemic states and response functions into $\V\s{ont}$.
Therefore, it is necessary that $m \leq \binom{k}{\lfloor k/2\rfloor}$.

Here, we give an outline proof of (b) and defer the full proof to Appendix~\ref{app:proof}.
Firstly, set $x:=\dim\spn\{\xi_i\}$, and $y:=\dim\spn\{\mu_i\}$. Also, after relabelling the preparation and outcomes in the COPE matrix, we may assume that $C_m$ has all-zero diagonal and strictly positive off-diagonal entries.

We prove $m\le F(x,y)$ by induction on $x+y$. The base cases are geometric. 
Consider the vectors of response-function values at each ontic point,
\begin{equation*}
(\xi_1(\lambda),\dots,\xi_m(\lambda)),
\qquad \lambda\in\Lambda.
\end{equation*} 
For $x=3$, these vectors lie in a two-dimensional affine slice of the simplex after an appropriate rescaling; hence, their convex hull is a polygon. The condition $0=(C_m)_{ii}=\langle \xi_i,\mu_i\rangle$ forces the support of $\mu_i$ to lie on the face where $\xi_i$ vanishes, while the positive off-diagonal entries imply that these supports form a Sperner family, with each support corresponding to a vertex or an edge of the polygon. This gives $\dim\spn\{\mu_i\}\ge m-1$, hence $m\le y+1=F(3,y)$. The case $y=3$ follows by the same argument with response functions and epistemic states interchanged.

For the induction step, consider an ontic coordinate $t$ and the
coordinate hyperplane $H_t:=\{z:z_t=0\}$. Then, the column indices of $C_m$ can be split according to whether the corresponding
epistemic states vanish on the ontic coordinate $t$:
$J_0:=\{j:\mu_j(t)=0\}$ and $J_1:=\{j:\mu_j(t)>0\}$. These give two disjoint
principal submatrices of $C_m$. In both cases, restriction to the hyperplane $H_t$ lowers one of the two spans
by one: on $J_0$ the epistemic states lie in $H_t$, giving
$|J_0|\le F(x,y-1)$, while on $J_1$ the diagonal-zero condition and
nonnegativity force the corresponding response functions to lie in $H_t$,
giving $|J_1|\le F(x-1,y)$. Therefore, by Pascal's identity,
\begin{equation*}
m=|J_0|+|J_1|
\le F(x,y-1)+F(x-1,y)
=F(x,y).
\end{equation*}
\end{proof}

\subsubsection{Proof that discrete-qubit theory is ontologically contextual}

Theorem~\ref{thm:Ontic_lowerbound}~(b) provides a link between the sparsity pattern of the COPE matrix and the minimum rank of its NMF factors. 
It remains to bound the rank of $C$ from above and show that it is lower than $\min\{\rank R,\rank P\}$, establish that the equirank condition in Eq.~\eqref{eq:OM_equirank} is impossible to satisfy.
Applying this technique to our example, we have the following.

\begin{theorem}\label{thm:qubit_contextuality}
    The discrete-qubit operational theory does not admit a noncontextual ontological model.
\end{theorem}
\begin{proof}
We assume that for every extremal preparation $\pp$ of the theory, its complementary one $\pp^\perp$ is also performed.
We also assume the set of measurements contains measurements that distinguish the two preparations perfectly, i.e., $p(1|\pp)=1$, $p(2|\pp)=0$, $p(1|\pp^\perp)=0$, and $p(2|\pp^\perp)=1$.
Note that $\pp$ is the unique preparation (up to operational equivalences) that never triggers the outcome $2$.
Similarly, $\pp^\perp$ is the unique preparation (up to operational equivalences) that never triggers the outcome $1$.
There are infinitely many such pairs of preparations and outcomes in $C\s{dq}$.
Then, by Theorem~\ref{thm:Ontic_lowerbound}~(b), the rank of at least one of the factors in any NMF of $C\s{dq}$ must be unbounded. 
Since the rank of $C\s{dq}$ is at most four, this implies a rank separation.
\end{proof}
In fact, by part (b) of Theorem~\ref{thm:Ontic_lowerbound}, the dimensionality of the span of response functions or epistemic states matrix of the smallest ontological model of $C\s{dq}$ grows logarithmically in the number of preparations, representing a logarithmic rank separation as shown in the following lemma.
\begin{lemma}
    Suppose $C_m$ is an $m\times m$ submatrix, with $m\ge3$, satisfying the hypotheses of
Theorem~\ref{thm:Ontic_lowerbound}~(b). For any ontological model
of $C_m$, let
\begin{equation*}
d_{\max}:=\max\{\dim\spn\{\xi_i\},\dim\spn\{\mu_i\}\}.
\end{equation*}
Then
\begin{equation*}
d_{\max}\ge
\max\left\{
3,
\left\lceil
2+\log_4\!\left(\frac{2m}{3}\right)
\right\rceil
\right\}.
\end{equation*}
In particular, at least one of the response-function span or the epistemic-state
span grows logarithmically in $m$.
\end{lemma}
\begin{proof}
By Theorem~\ref{thm:Ontic_lowerbound}~(b),
\begin{equation*}
m\le F(\dim\spn\{\xi_i\},\dim\spn\{\mu_i\}).
\end{equation*}
Since $F$ is symmetric in its arguments and nondecreasing,
\begin{equation*}
\begin{split}
m&\le F(d_{\max},d_{\max})\\
&=
\binom{2d_{\max}-4}{d_{\max}-2}
+
2\binom{2d_{\max}-6}{d_{\max}-3}\\
&\le
\frac{3}{2}4^{d_{\max}-2}.
\end{split}
\end{equation*}
Solving for $d_{\max}$ gives the result.
\end{proof}
We note that our qubit example gives another proof of the quantum excess-baggage theorem~\cite{Hardy2004,Montina2006,Montina2008,Jennings2016}.
It follows that, in general, the contextuality of operational theory implies ontological excess baggage.
Note, however, that the converse is not necessarily true because there are operational theories, such as Spekkens' toy theory, that accommodate a noncontextual ontological model only with an overhead in the size of their parameter space, i.e., with an excess baggage.

\section{Convexity preservation} \label{sec:Conv}

We showed in the previous section how to construct five different types of models, (i) preGPT, (ii) GPT, (iii) quasiprobabilistic, (iv) ontological, and (v) noncontextual ontological (subject to existence), for given operational theories.
We have already seen that the relationship between linearly isomorphic models is clear.
For example, a GPT is obtained from a preGPT by a linear projection.
Also, a quasiprobabilistic model or, more generally, any GPT model is obtained from a GPT by a linear transformation.
Here, our goal is to pinpoint a more general feasibility condition: under what conditions can a GPT be mapped to an ontological model?

The standard approach to map a GPT to an ontological model introduces two linear \textit{functions} $\tilde{\mu}:\St\s{GPT} \to \mathcal{P}(\Lambda)$ and $\tilde{\xi}:\EE\s{GPT} \to \mathcal{C}_b(\Lambda)$ where $\mathcal{P}(\Lambda)$ and $\mathcal{C}_b(\Lambda)$ are the sets of all probability measures and all bounded real-valued function on $\Lambda$, respectively.
The images of $\tilde{\mu}$ and $\tilde{\xi}$ are epistemic states and response functions, respectively.
Thanks to our linear algebraic framework expressing models as matrix factorizations of the COPE matrix $C$, $\tilde{\mu}$ and $\tilde{\xi}$ correspond to linear transformations $T_{\tilde{\xi}}$ and $T_{\tilde{\mu}}$ such that $C=MS=(MT_{\tilde{\xi}}) (T_{\tilde{\mu}}S):=RP$.
Linear maps cannot increase rank.
Furthermore, the probabilistic structure must be preserved, that is, $T_{\tilde{\xi}}$ and $T_{\tilde{\mu}}$ must not reduce the ranks of $M$ and $S$.
Hence, they must preserve the GPT equirank condition of Eq.~\eqref{eq:GPT_equirank}, and we have
\begin{equation}
    \rank C = \rank M = \rank S = \rank R = \rank P.
\end{equation}
Crucially, any such equirank-preserving linear transformations yield a noncontextual ontological model (Eq.~\eqref{eq:OM_equirank}). 
Thus, in this picture, a GPT can be mapped to an ontological model if and only if the operational theory admits a \textit{noncontextual} ontological model~\cite{Schmid2021}.

In this approach, for example, the GPT model for the boxworld in Eq.~\eqref{eq:BW_GPT} cannot be mapped to an ontological model, even though the operational theory itself admits an ontological model as presented in Eq.~\eqref{eq:BW_OM}. 
This is justified on the grounds that quotienting the operational theory to obtain the GPT removes the contextual structure which is present in the ontological model, causing the mapping to fail~\cite{Schmid2021}.

We now argue that requiring $\tilde{\mu}$ and $\tilde{\xi}$ to be \textit{simultaneously} linear \textit{functions} is unnecessarily restrictive, thus providing an alternative perspective.
In particular, we emphasize that the usual justification that $\tilde{\mu}$ and $\tilde{\xi}$ must preserve convexity does not imply they are \textit{single-valued} functions.
Instead, we can allow linear \textit{set-valued} functions\footnote{Such functions are also known under the name \textit{multivalued} functions.} which preserve convexity between equivalence classes of objects~\cite{Dacic1969}.
\begin{definition}\label{def:set_val_fun}
    A set-valued function between two vector spaces $\V$ and $\mathcal{W}$, $f:\V\leadsto\mathcal{W}$, is one such that $f(v)\subseteq \mathcal{W}$ for every $v\in\V$.
    Furthermore, $f$ is said to be linear if and only if,
    \begin{enumerate}[label=(\alph{enumi})]
        \item $w_1\in f(v_1)$ and $w_2\in f(v_2)$ imply that $w_1+w_2 \in f(v_1+v_2)$,
        \item $w\in f(v)$ and $\alpha$ a scalar imply that $\alpha w \in f(\alpha v)$.
    \end{enumerate}
    Finally, $f$ is said to be convex if it is linear and $w_1\in f(v_1)$, $w_2\in f(v_2)$, and $p\in[0,1]$ imply that $p w_1+(1-p)w_2 \in f(pv_1+(1-p)v_2)$.
\end{definition}
\noindent Now, let $\tilde{\mu}:\St\s{GPT}\leadsto\PP(\Lambda)$ with $\varrho \mapsto [\mu_\varrho]$ be a set-valued function.
Here, $[\mu_\varrho]$ denotes an equivalence class of epistemic states such that $\inprod{\xi}{\mu}=\inprod{\xi}{\mu'}$ for any pair of epistemic distributions $\mu,\mu'\in[\mu_\varrho]$ and all $\xi\in\RF$\footnote{Note that, unless $[\mu_\varrho]$ is a singleton, such an ontological model is contextual because it is not tomographically complete.}.
If $\varrho_1$, $\varrho_2$, and $\varrho_3$ are GPT states such that $\varrho_3=p\varrho_1+(1-p)\varrho_2$, convexity preservation for states requires only that $p\mu_1+(1-p)\mu_2\in [\mu_3]$.
This is precisely the convexity criterion for set-valued functions in Definition~\ref{def:set_val_fun} above.
By contrast, the standard function approach forces $[\mu_\varrho]$ to be a singleton so that $\mu_3=p\mu_1+(1-p)\mu_2$, which means the map $\tilde{\mu}$ not only preserves convexity, but also preserves tomographic completeness of the GPT.
A similar argument can be applied to conclude that a convexity-preserving $\tilde{\xi}$ is most generally described by a set-valued function.

As we will constructively show now, relaxing (at least) one of $\tilde{\mu}$ and $\tilde{\xi}$ to be a set-valued function enables us to obtain ontological models for all GPTs.
Let us define $\tilde{\xi}$ as
\begin{equation}
    e_i\mapsto \xi_i, \quad (\xi_i)_j:=\inprod{e_i}{\rho_j}~\forall \rho_j\in{\rm ex}\St\s{GPT}.
\end{equation}
This is equivalent to transforming the GPT effects into ontological response functions using the extremal GPT states so that the image of $\tilde{\xi}$ is simply the row space of the COPE matrix.
Hence, $\tilde{\xi}$ is a standard function (linear transformation).
Then, $\tilde{\mu}$ will be a well-defined \textit{set-valued} function given by 
\begin{equation}
\begin{split}
    &\rho_i\mapsto[\mu_i],\\
    &[\mu_i]:=\{\mu|\mu\geq 0 \land \inprod{\tilde{\xi}(e_j)}{\mu}=\inprod{e_j}{\rho_i}~ \forall e_j\in{\rm ex}\EE\s{GPT} \}.
\end{split}
\end{equation}
Note that this function is single-valued over the extremal states, i.e., $\rho_i\mapsto (\delta_{ij})_j$ for $\rho_i\in{\rm ex}\St\s{GPT}$, and results in set-valued outputs over mixed states only.

The construction above is valid for any GPT and is equivalent to the trivial ontological model~\cite{Beltrametti1995}.
We also note that the use of set-valued functions does not change the fact that, whenever a noncontextual ontological model exists, there exist single-valued linear functions from the GPT to the noncontextual ontological model.
We summarize the above discussions in the following corollary.

\begin{cor} \label{cor:GPT_OM_map}
    Every GPT admits an ontological model, where at least one of $\tilde{\xi}$ and $\tilde{\mu}$ is a linear single-valued function while the other one is a linear set-valued function.
    The ontological model is noncontextual if and only if both $\tilde{\xi}$ and $\tilde{\mu}$ are linear single-valued functions.
\end{cor}

\section{Discussion and Conclusion}\label{sec:concl}

In this work, we introduced a novel framework for analyzing operational theories and their nonclassical properties, built from a statistics-first perspective.
At its core lies the matrix of conditional outcome probabilities of events (COPE), which captures the complete probabilistic structure of the operational theory. 
By reinterpreting different classes of abstract physical models, namely preGPTs, GPTs, quasiprobabilistic, ontological, and noncontextual ontological models, as distinct factorizations of the COPE matrix, we have unified them within a single, transparent linear-algebraic structure.

This perspective clarifies the interrelationship between the probabilistic structure of an operational theory, the mathematical constraints on its matrix factorization, and the resulting model characteristics.
We showed that the assumption of tomographic completeness is equivalent to an equirank condition on the COPE matrix factors for all models, including GPTs and ontological models. 
While GPTs and quasiprobabilistic models always satisfy this condition, its fulfillment for ontological models is not guaranteed.
Such failure, namely \emph{rank‑separation}, is both necessary and sufficient for contextuality. Thus, an operational theory admits a noncontextual ontological model if and only if its COPE matrix allows for an equirank nonnegative matrix factorization (ENMF).
This powerful, GPT-agnostic criterion enables the identification of contextuality directly from observable statistics, without recourse to an underlying abstract model or an explicit analysis of operational equivalences.

We have illustrated our framework and results by applying them to canonical exemplary operational theories.
Spekkens’ toy theory served to clarify the probabilistic structure of an operational theory and to demonstrate the construction of all five model types.
The rank‑separation criterion was then applied in two contrasting proofs: a geometric argument establishing the contextuality of boxworld, and an ontic‑space dimensional‑counting argument for the qubit theory. 
The latter also yields a fresh perspective on Hardy’s quantum excess‑baggage theorem, linking it directly to COPE structure and contextuality.

Finally, by reframing contextuality as a problem in matrix theory and geometry, our framework not only unifies disparate model classes but also lays the groundwork for future work. 
These include developing quantitative measures of contextuality based on rank separation, extending the framework to accommodate continuous-variable models, and establishing a systematic classification of nonclassicality. 
This approach promises to significantly advance the foundations and applications of nonclassical resources.
~\\

\acknowledgments{FS and MD gratefully acknowledge the financial support from the Engineering and Physical Sciences Research Council (EPSRC) through the Hub in Quantum Computing and Simulation grant (EP/T001062/1). TY acknowledges the support through the Quantum Computing Studentship funded by Royal Holloway, University of London.}

\bibliography{quantum}
\bibliographystyle{apsrev4-1}

\begin{widetext}

\appendix

\setcounter{equation}{0}
\renewcommand{\theequation}{\thesection.\arabic{equation}}

\newpage

\section{Faithfulness and relative tomographic completeness}\label{app:Rel_TC}

The possibility of extending or restricting operational theories by enlarging or selecting subsets of their preparation and measurement sets raises subtle questions about how contextuality behaves under such extensions.
For example, it is often stated that fragments of a noncontextual GPT are noncontextual.
The justification is that (i) every state and effect of the fragment is a state and effect in the true GPT; 
(ii) Thus, the operational identities that hold in the fragment are just a subset of those holding in the full GPT;
(iii) Restricting the noncontextual ontological model of the true GPT so that it automatically provides a valid ontic representation for the fragment will preserve the operational identities.
Hence, it gives a valid noncontextual ontological model for the fragment.
In this section, we aim to demonstrate that this justification is based on the \textit{faithfulness} assumption~\ref{ass:faithful}.

Let us show via explicit examples that without the faithfulness assumption, a fragment COPE matrix $C$ may or may not be contextual, regardless of whether the parent theory $D$ is contextual or noncontextual; the outcome depends on the specific form of the restriction.
\begin{enumerate}[label=\roman*.]
    \item Every operational theory can trivially be restricted to $\M\s{C}\subseteq\M\s{D}$ and $\PP\s{C}\subseteq\PP\s{D}$ where $\M\s{C}$ and $\PP\s{C}$ are singletons. Without the faithfulness assumption, the corresponding COPE matrix $C$ will be noncontextual, because there is a unique ontological representation for the preparation and the measurement.
    \item A noncontextual COPE matrix can also be restricted to a contextual one.
    For instance, restricting Spekkens’ toy theory to
\begin{gather}
C := 
\begin{pmatrix}
1 & 0 & \frac{1}{2} & \frac{1}{2} \\
0 & 1 & \frac{1}{2} & \frac{1}{2} \\
\frac{1}{2} & \frac{1}{2} & 1 & 0 \\
\frac{1}{2} & \frac{1}{2} & 0 & 1 
\end{pmatrix}
\end{gather}
yields a contextual COPE matrix when faithfulness is not assumed. 
    \item Interestingly, there are (noncontextual) COPE matrices such that any of their restrictions are noncontextcual. An example is $D=\Ident$.
    \item A contextual COPE matrix may be restricted to a noncontextual one. 
    For example, the qubit operational theory restricted to Spekkens’ toy theory becomes noncontextual.
    Note that, this is only true if one does not assume faithfulness.
    \item In contrast, restricting the qubit theory to preparations and measurements confined to the equator of the Bloch sphere preserves contextuality (with or without faithfulness).
\end{enumerate}
We thus observe that, when restricting operational theories, preserving noncontextuality is not automatic without additional assumptions.
In other words, none of the justifications (i)-(iii) are true without the faithfulness assumption.

In Ref.~\cite{schmid2024shadows}, \textit{relative tomographic completeness} for GPT fragments was introduced, in which a subset of GPT states and a subset of GPT effects are tomographically complete with respect to one another.
This can be formally stated as follows.
\begin{definition} \textbf{(Relative Tomographic Completeness).} \label{def:Rel_TC_MS}
    The strict subsets ${\rm ex}\EE_{\rm F}$ and $ex\St_{\rm F}$ of a GPT's extremal effects and states, ${\rm ex}\EE_{\rm GPT}$ and $ex\St_{\rm GPT}$, respectively, are relatively tomographically complete measurement and states if and only if:
    \begin{enumerate}[label=\roman*.]
        \item For any two states $\rho_i,\rho_j \in \St_{\rm F}$, $\rho_i = \rho_j$ if and only if $\inprod{\rho_i}{e} = \inprod{\rho_j}{e}$ for all effects $e \in {\rm ex}\EE_{\rm F}$.
        \item For any two effects $e_i,e_j \in \EE_{\rm F}$, $e_i = e_j$ if and only if $\inprod{\rho}{e_i} = \inprod{\rho}{e_j}$ for all states $\rho \in {\rm ex}\St_{\rm F}$.
    \end{enumerate}
\end{definition}
\noindent The relative tomographic completeness asserts that, once \textit{specific subsets of state and effect vectors} are designated as admissible, the observed statistics suffice to uniquely identify the corresponding vectors \textit{within those subsets.} 
Importantly, any state in the parent GPT, $\psi\in \St$, which agrees with the states in $\St\s{F}$ on restricted effects ${\rm ex}\EE\s{F}$ but differs on the full set of measurements $\EE$ is, \textit{by assumption}, excluded from the restricted set $\St\s{F}$.
Similarly, any effect in the parent GPT which could not be separated from the effects in the restricted set, $\EE\s{F}$, using the restricted states is, \textit{by assumption}, excluded from the restricted set $\EE\s{F}$.
Relative tomographic completeness thus entails two implicit assumptions:
First, a parent operational theory exists and states and effects of the fragment theory share the representation with the parent theory--this is faithfulness (Assumption~\ref{ass:faithful}).
Second, every state and effect which could have been possibly mistaken on the restricted sets of effects and states are excluded by assumption.
It thus follows that the relative tomographic completeness is not simply a weaker form of tomographic completeness, but a different assumption altogether.
An assumption that trades global completeness for two strong assumptions.

The above assumption can also be expressed in the language of COPE matrices.
Let $D$ denote the COPE matrix of the operational theory, and let $C$ be a restricted COPE matrix for subsets $\M\s{C}$ and $\PP\s{C}$ of all measurements and preparations, respectively, as before.
Relative tomography then implies that if two columns $C_{:i}$ and $C_{:j}$ of $C$ are identical, their corresponding columns in the full matrix $D$, $D_{:i}$ and $D_{:j}$, are also identical.
It is important to emphasize, however, that the converse need not hold: a column $D_{:k}$ may differ from $C_{:i}$ and $C_{:j}$ in the full matrix, while still agreeing with them on all rows indexed by outcomes of the restricted measurements $\M\s{C}$.
Thus, the ability to uniquely identify columns of $C$ rests on the promise that any column of $D$ which could be indistinguishable under the restricted measurements but distinguishable under the full set of measurements $\M\s{D}$ is, by assumption, excluded from the restricted theory.
A similar argument applies to measurement effects.

\section{Extended boxworld}\label{app:exBW}

We have seen how different types of models arise from factorizations of the COPE matrix under different requirements.
In this example, we would like to clarify the points made in Sec.~\ref{sec:IRC_rep}.
To achieve this, consider the following COPE matrix for a fictitious operational theory which we refer to as the \textit{extended boxworld}:
\begin{equation}\label{eq}
C\s{EBW} := 
\begin{pmatrix}
0 & 0 & 0 & 0 & 1 & 1\\
1 & 0 & 0 & 1 & 0 & 0\\
0 & 1 & 1 & 0 & 0 & 0\\
0 & 0 & 0 & 0 & 1 & 1\\
1 & 0 & 1 & 0 & 0 & 0\\
0 & 1 & 0 & 1 & 0 & 0\\
\end{pmatrix}.
\end{equation}
We can easily identify a $4\times 4$ submatrix of $C\s{EBW}$ which is identical to the boxworld in Eq.~\eqref{eq:C_BW}.
However, in this world, there are two measurements such that their first outcomes (the first and fourth rows of $C\s{EBW}$) are operationally indistinguishable.
Furthermore, the fifth and sixth preparations are operationally equivalent.
We begin with the SVD of the COPE,
\begin{equation}
\begin{split}
& U\s{EBW}=
    \begin{pmatrix}
\frac{1}{\sqrt{2}} & 0 & 0 & 0 & 0 & \frac{1}{\sqrt{2}} \\
0 & \frac{1}{2} & \frac{1}{2} & -\frac{1}{2} & \frac{1}{2} & 0 \\
0 & \frac{1}{2} & -\frac{1}{2} & \frac{1}{2} & \frac{1}{2} & 0 \\
\frac{1}{\sqrt{2}} & 0 & 0 & 0 & 0 & \frac{1}{\sqrt{2}} \\
0 & \frac{1}{2} & -\frac{1}{2} & -\frac{1}{2} & \frac{1}{2} & 0 \\
0 & \frac{1}{2} & \frac{1}{2} & \frac{1}{2} & \frac{1}{2} & 0
\end{pmatrix},\\
& \Sigma\s{EBW}=
\begin{pmatrix}
2 & 0 & 0 & 0 & 0 & 0 \\
0 & 2 & 0 & 0 & 0 & 0 \\
0 & 0 & \sqrt{2} & 0 & 0 & 0 \\
0 & 0 & 0 & \sqrt{2} & 0 & 0 \\
0 & 0 & 0 & 0 & 0 & 0 \\
0 & 0 & 0 & 0 & 0 & 0
\end{pmatrix},\\
& V\s{EBW}=
\begin{pmatrix}
0 & 0 & 0 & 0 & \frac{1}{\sqrt{2}} & \frac{1}{\sqrt{2}} \\
\frac{1}{2} & \frac{1}{2} & \frac{1}{2} & \frac{1}{2} & 0 & 0 \\
0 & 0 & -\frac{1}{\sqrt{2}} & \frac{1}{\sqrt{2}} & 0 & 0 \\
-\frac{1}{\sqrt{2}} & \frac{1}{\sqrt{2}} & 0 & 0 & 0 & 0 \\
0 & 0 & 0 & 0 & -\frac{1}{\sqrt{2}} & \frac{1}{\sqrt{2}} \\
-\frac{1}{2} & -\frac{1}{2} & \frac{1}{2} & \frac{1}{2} & 0 & 0
\end{pmatrix}.    
\end{split}
\end{equation}
This gives the states
\begin{equation}
    B\s{EBW}:=\Sigma\s{EBW}V\s{EBW}=
    \begin{pmatrix}
0 & 0 & 0 & 0 & \sqrt{2} & \sqrt{2} \\
1 & 1 & 1 & 1 & 0 & 0 \\
0 & 0 & -1 & 1 & 0 & 0 \\
-1 & 1 & 0 & 0 & 0 & 0 \\
0 & 0 & 0 & 0 & 0 & 0 \\
0 & 0 & 0 & 0 & 0 & 0
\end{pmatrix}
\end{equation}
and the effects $U\s{EBW}$ for a preGPT of this operational theory such that $C\s{EBW}=U\s{EBW}B\s{EBW}$.
The unit effect of this model is given by $u\s{EBW}=(1/\sqrt{2},1,0,0,1,1/\sqrt{2})$.
We observe that in this preGPT operationally equivalent outcomes in rows one and four are represented by identical vectors in $U\s{EBW}$.
Furthermore, operationally equivalent preparations five and six have been mapped to identical state vectors as expected.

However, one might ask whether they can modify this preGPT such that it tracks the contexts.
In general, if there are $n$ equivalent outcomes and $m$ equivalent preparations to track, this can be naively done by embedding the model in a vector space with $n+m$ extra dimensions in such a way that equivalent outcomes and equivalent preparations live in orthogonal subspaces.
The latter ensures that the statistics are reproduced correctly.
We now show that there might exist clever ways to reduce this dimensional overhead.

In particular, in our extended boxworld example, we do not need to add any extra dimensions:
We have six dimensions to accommodate a rank-four model, leaving two unused dimensions to exploit for distinguishing the two outcomes and the two preparations.
We thus modify the last two columns of $U\s{EBW}$ and the last two rows of $B\s{EBW}$ to obtain
\begin{equation}\label{eq:EBW_preGPT_2}
    \begin{split}
        & U'\s{EBW}=
    \begin{pmatrix}
\frac{1}{\sqrt{2}} & 0 & 0 & 0 & 0 & 0 \\
0 & \frac{1}{2} & \frac{1}{2} & -\frac{1}{2} & 0 & 0 \\
0 & \frac{1}{2} & -\frac{1}{2} & \frac{1}{2} & 0 & 1 \\
\frac{1}{\sqrt{2}} & 0 & 0 & 0 & 0 & 1 \\
0 & \frac{1}{2} & -\frac{1}{2} & -\frac{1}{2} & 0 & 0 \\
0 & \frac{1}{2} & \frac{1}{2} & \frac{1}{2} & 0 & 0
\end{pmatrix},\\
        & B'\s{EBW}=
    \begin{pmatrix}
0 & 0 & 0 & 0 & \sqrt{2} & \sqrt{2} \\
1 & 1 & 1 & 1 & 0 & 0 \\
0 & 0 & -1 & 1 & 0 & 0 \\
-1 & 1 & 0 & 0 & 0 & 0 \\
0 & 0 & 0 & 0 & 1 & 0 \\
0 & 0 & 0 & 0 & 0 & 0
\end{pmatrix}.
    \end{split}
\end{equation}
The unit vector of the modified preGPT is $u'\s{EBW}=(1/\sqrt{2},1,0,0,0,1)$.
We note that both preGPT models given above violate the preGPT tomographic completeness~\ref{def:TC} and thus do not meet the broad noncontextuality assumption.
This means that the statistics in the COPE matrix alone are insufficient to uniquely identify the states and effects of these models.
This was evidenced by our ability to freely modify two rows and two columns of the effects and states matrices, respectively.

A GPT for this operational theory can be constructed by removing the extra dimensions from $U\s{EBW}$ and $B\s{EBW}$ to obtain
\begin{equation}\label{eq:EBW_GPT}
        \begin{split}
        & M\s{EBW}=
    \begin{pmatrix}
\frac{1}{\sqrt{2}} & 0 & 0 & 0  \\
0 & \frac{1}{2} & \frac{1}{2} & -\frac{1}{2}  \\
0 & \frac{1}{2} & -\frac{1}{2} & \frac{1}{2}  \\
\frac{1}{\sqrt{2}} & 0 & 0 & 0 \\
0 & \frac{1}{2} & -\frac{1}{2} & -\frac{1}{2} \\
0 & \frac{1}{2} & \frac{1}{2} & \frac{1}{2} 
\end{pmatrix},\\
        & S\s{EBW}=
    \begin{pmatrix}
0 & 0 & 0 & 0 & \sqrt{2} & \sqrt{2} \\
1 & 1 & 1 & 1 & 0 & 0 \\
0 & 0 & -1 & 1 & 0 & 0 \\
-1 & 1 & 0 & 0 & 0 & 0 
\end{pmatrix}.
    \end{split}
\end{equation}
As promised, operationally equivalent extremal outcomes and preparations are represented by identical vectors in the model, ensuring its tomographic completeness.

To obtain a quasiprobabilistic model, we choose the submatrix of $S\s{EBW}$,
%\
\begin{equation}
    T:=
    \begin{pmatrix}
0 & 0 & 0 & \sqrt{2} \\
1 & 1 & 1 & 0  \\
0 & 0 & -1 & 0  \\
-1 & 1 & 0 & 0  
\end{pmatrix},
\end{equation}
so that
\begin{equation} \label{eq:EBW_Quasi}
        \begin{split}
        & M'\s{EBW}=M\s{EBW}T=
    \begin{pmatrix}
0 & 0 & 0 & 1 \\
1 & 0 & 0 & 0 \\
0 & 1 & 1 & 0 \\
0 & 0 & 0 & 1 \\
1 & 0 & 1 & 0 \\
0 & 1 & 0 & 0
\end{pmatrix},\\
        & S'\s{EBW}=T^{-1}S\s{EBW}=
    \begin{pmatrix}
1 & 0 & 0 & 1 & 0 & 0 \\
0 & 1 & 0 & 1 & 0 & 0 \\
0 & 0 & 1 & -1 & 0 & 0 \\
0 & 0 & 0 & 0 & 1 & 1
\end{pmatrix}.
    \end{split}
\end{equation}
The model is quasiprobabilistic as evidenced by the negativity in the states, and its unit vector is $u\s{EBW}=(1,1,1,1)$ as required.

We now show that the extended boxworld theory does not admit a noncontextual ontological model.
The argument is similar to that in the boxworld of the previous section, but here in six dimensions.
Every column of (the column-stochastic form of) $C\s{EBW}$ is a vertex of a polytope $\mathbf {C}\s{EBW}$ in the standard 5-simplex $\Delta_6$.
The zeros in the columns of $C\s{EBW}$ indicate that the vertices of $\mathbf{C}\s{EBW}$ form a pentachoron whose vertices lie on the edges of $\Delta_6$.
For $C\s{EBW}$ to admit a noncontextual ontological model it is necessary that there exists another polytope in $\Delta_6$ living in the same subspace as $\mathbf{C}\s{EBW}$ containing it.
Since every vertex of $C\s{EBW}$ lies on an edge (1-face) of $\Delta_6$, and it must be written as a convex combination of some vertices of the presumed intermediate polytope, some vertices of the latter must coincide with those of $\Delta_6$.
Now, none of the vertices of $\Delta_6$ lie in the subspace spanned by $C\s{EBW}$, implying that the desired intermediate simplex does not exist.

An ontological model for the extended boxworld theory satisfying $\rank{W\s{EBW}}=\rank{C\s{EBW}}$ is given by
\begin{equation}\label{eq:EBW_OM_1}
       W\s{EBW}=
\begin{pmatrix}
0 & 0 & 0 & 0 & 1 \\
1 & 0 & 0 & 1 & 0 \\
0 & 1 & 1 & 0 & 0 \\
0 & 0 & 0 & 0 & 1 \\
1 & 0 & 1 & 0 & 0 \\
0 & 1 & 0 & 1 & 0 \\
\end{pmatrix},
H\s{EBW} = 
\begin{pmatrix}
1 & 0 & 0 & 0 & 0 & 0\\
0 & 1 & 0 & 0 & 0 & 0 \\
0 & 0 & 1 & 0 & 0 & 0 \\
0 & 0 & 0 & 1 & 0 & 0 \\
0 & 0 & 0 & 0 & 1 & 1 
\end{pmatrix}.
\end{equation}
In this model, the representations of \textit{extremal} equivalent preparations and outcomes are identical and unique.
Nevertheless, it is contextual because $\rank{H\s{EBW}}>\rank{W\s{EBW}}$.
As argued in Sec.~\ref{sec:IRC_rep}, the contextuality arises from the fact that the uniqueness of representations cannot be preserved for convexly dependent preparations and outcomes.
For example, the balanced mixture of the first and second ontic states, $\mu_{12}=(1/2,1/2,0,0,0)\trs$, is operationally indistinguishable from the balanced mixture of the third and fourth states, $\mu_{34}=(0,0,1/2,1/2,0,0)$,
\begin{equation}
    W\s{EBW}\mu_{12} = 
    \begin{pmatrix}
        0\\
        \frac{1}{2}\\
        \frac{1}{2}\\
        0\\
        \frac{1}{2}\\
        \frac{1}{2}
    \end{pmatrix}
    = W\s{EBW\mu_{34}}.
\end{equation}
As discussed in Sec.~\ref{sec:IRC_rep}, we could assume that the extremal rows and columns of $C\s{EBW}$ are quotiented with respect to operational equivalences up to repetitions of equivalent outcomes in \textit{distinct} measurements to get
\begin{equation} \label{eq:EBW_COPE_quotiented}
 C\s{EBW}^{\rm quotient} := 
\begin{pmatrix}
0 & 0 & 0 & 0 & 1 \\
1 & 0 & 0 & 1 & 0 \\
0 & 1 & 1 & 0 & 0 \\
0 & 0 & 0 & 0 & 1 \\
1 & 0 & 1 & 0 & 0 \\
0 & 1 & 0 & 1 & 0 \\
\end{pmatrix}.
\end{equation}
Note that the first and fourth outcomes are not quotiented out because they belong to two distinct measurements.
The models of $C\s{EBW}^{\rm quotient}$ simply amount to removing the repeated rows and columns from the effect and state matrices in Eqs.~\eqref{eq:EBW_GPT},~\eqref{eq:EBW_Quasi}, and~\eqref{eq:EBW_OM_1}.

We can compare the above models with the trivial ontological model of the unquotiented COPE matrix $C\s{EBW}$,
\begin{equation}
       W'\s{EBW}=
\begin{pmatrix}
0 & 0 & 0 & 0 & 1 & 1\\
1 & 0 & 0 & 1 & 0 & 0\\
0 & 1 & 1 & 0 & 0 & 0\\
0 & 0 & 0 & 0 & 1 & 1\\
1 & 0 & 1 & 0 & 0 & 0\\
0 & 1 & 0 & 1 & 0 & 0\\
\end{pmatrix},
H'\s{EBW} = 
\begin{pmatrix}
1 & 0 & 0 & 0 & 0 & 0\\
0 & 1 & 0 & 0 & 0 & 0 \\
0 & 0 & 1 & 0 & 0 & 0 \\
0 & 0 & 0 & 1 & 0 & 0 \\
0 & 0 & 0 & 0 & 1 & 0 \\
0 & 0 & 0 & 0 & 0 & 1
\end{pmatrix}.
\end{equation}
which tracks all preparations but not outcomes, and the enlarged model,
\begin{equation}
\begin{split}
      & W''\s{EBW}=
\begin{pmatrix}
0 & 0 & 0 & 0 & 1 & 1 & 0\\
1 & 0 & 0 & 1 & 0 & 0 & 0\\
0 & 1 & 1 & 0 & 0 & 0 & 1\\
0 & 0 & 0 & 0 & 1 & 1 & 1\\
1 & 0 & 1 & 0 & 0 & 0 & 0\\
0 & 1 & 0 & 1 & 0 & 0 & 0\\
\end{pmatrix},\\
& H''\s{EBW} = 
\begin{pmatrix}
1 & 0 & 0 & 0 & 0 & 0\\
0 & 1 & 0 & 0 & 0 & 0 \\
0 & 0 & 1 & 0 & 0 & 0 \\
0 & 0 & 0 & 1 & 0 & 0 \\
0 & 0 & 0 & 0 & 1 & 0 \\
0 & 0 & 0 & 0 & 0 & 1 \\
0 & 0 & 0 & 0 & 0 & 0
\end{pmatrix},
\end{split}
\end{equation}
which also tracks repeated outcomes in exchange for an additional ontic dimension.

\section{Proof of Theorem 2~(b)} \label{app:proof}
\begin{proof}
After relabelling the rows and columns, the submatrix $C_m$ satisfies
\begin{equation*}
(C_m)_{ii}=0,
\qquad
(C_m)_{ij}>0 \quad(i\neq j).
\end{equation*}
We prove by induction on $x+y$ the following claim: for every
$C_n\in\mathbb R_{\ge0}^{n\times n}$ satisfying
\begin{equation*}
(C_n)_{ii}=0,
\qquad
(C_n)_{ij}>0 \quad(i\neq j),
\end{equation*}
and every nonnegative factorization
\begin{equation*}
C_n=RP,
\qquad
\rank(R)\le x,
\qquad
\rank(P)\le y,
\end{equation*}
one has
\begin{equation*}
n\le F(x,y),
\end{equation*}
where
\begin{equation*}
F(x,y):=
\binom{x+y-4}{x-2}
+
2\binom{x+y-6}{x-3}.
\end{equation*}
Here the rows of $R$ are the response functions $\xi_i$, and the columns of
$P$ are the epistemic states $\mu_i$. Applying the claim to $C_m$, with
$x=\dim\spn\{\xi_i\}$ and $y=\dim\spn\{\mu_i\}$, gives the desired result.

If $n\le2$, the claim is immediate. For $n\ge3$, every $3\times3$ principal
submatrix has the form
\begin{equation*}
\begin{pmatrix}
0&a&b\\
c&0&d\\
e&f&0
\end{pmatrix}
\qquad
(a,b,c,d,e,f>0),
\end{equation*}
whose determinant is $ade+bcf>0$. Hence $\rank(C_n)\ge3$, and the nontrivial
cases have $x,y\ge3$. Also,
\begin{equation*}
F(3,y)=y+1,
\qquad
F(x,3)=x+1,
\end{equation*}
and, for $x,y\ge4$, Pascal's identity gives
\begin{equation}\label{eq:F-rec}
F(x,y)=F(x,y-1)+F(x-1,y).
\end{equation}

Consider first the base case $x=3$. Since $\rank(C_n)\ge3$ and
$\rank(R)\le3$, we have $\rank(R)=3$. After deleting unused ontic coordinates
and rescaling, we may assume that each column $R_{:\lambda}$ is stochastic.
The convex hull
\begin{equation*}
H:=\conv\{R_{:\lambda}:\lambda\in\Lambda\}
\end{equation*}
is therefore a polygon in an affine two-dimensional slice of the simplex.
Replacing the columns $R_{:\lambda}$ by the vertices $h_1,\dots,h_\ell$ of
$H$, and absorbing the corresponding convex coefficients into $P$, gives
another nonnegative factorization
\begin{equation*}
C_n=\widetilde R\,\widetilde P,
\qquad
\rank(\widetilde P)\le y.
\end{equation*}

For each $i$, let
\begin{equation*}
G_i:=H\cap\{z:z_i=0\},
\qquad
A_i:=\{a:(\widetilde P)_{ai}>0\}.
\end{equation*}
The set $A_i$ is the set of new ontic coordinates, equivalently vertices of
$H$, assigned positive weight by the $i$th epistemic state. Since
\begin{equation*}
0=(C_n)_{ii}
=
\sum_{a=1}^{\ell}(h_a)_i(\widetilde P)_{ai}
\end{equation*}
and all summands are nonnegative, every $a\in A_i$ satisfies $(h_a)_i=0$.
Thus the vertices assigned positive weight by the $i$th epistemic state lie on
the face $G_i$. This face is nonempty and proper. Hence, since $H$ is a
polygon, $A_i$ is either a single vertex or the two endpoints of an edge.
Moreover, if $A_j\subseteq A_i$ for $j\neq i$, then every vertex assigned
positive weight by the $j$th epistemic state lies on the face where $\xi_i$
vanishes, which implies
\begin{equation*}
(C_n)_{ij}=0,
\end{equation*}
contradicting the strict positivity of the off-diagonal entries. Therefore the
sets $A_i$ form a Sperner family of vertices and edges of a cycle.

It follows that, after reordering the new ontic coordinates and epistemic
states, $\widetilde P$ decomposes into positive $1\times1$ blocks, bidiagonal
path blocks, and at most one cyclic bidiagonal block. The path blocks have full
column rank. A cyclic block can lose at most one dimension, because any linear
relation among its columns determines all coefficients from one coefficient.
Therefore
\begin{equation*}
\rank(\widetilde P)\ge n-1.
\end{equation*}
Since $\rank(\widetilde P)\le y$, we obtain
\begin{equation*}
n\le y+1=F(3,y).
\end{equation*}
The case $y=3$ follows by applying the same argument to
\begin{equation*}
C_n^\top=P^\top R^\top,
\end{equation*}
which interchanges response functions and epistemic states. Thus
\begin{equation*}
n\le x+1=F(x,3).
\end{equation*}

Now assume $x,y\ge4$, and suppose the claim holds for smaller values of
$x+y$. Delete unused ontic coordinates, so that every ontic coordinate is used
by some response function and some epistemic state. Choose an ontic coordinate
$t$, and let
\begin{equation*}
H_t:=\{z:z_t=0\}.
\end{equation*}
Split the labels of the epistemic states into
\begin{equation*}
J_1:=\{j:\mu_j(t)>0\},
\qquad
J_0:=\{j:\mu_j(t)=0\}.
\end{equation*}
The principal submatrices $(C_n)_{J_0,J_0}$ and $(C_n)_{J_1,J_1}$ again have
zero diagonal and strictly positive off-diagonal entries.

On $J_0$, the epistemic states $\{\mu_j:j\in J_0\}$ lie in $H_t$. Since some
epistemic state has nonzero $t$th coordinate, this lowers the epistemic-state
span by one:
\begin{equation*}
\dim\spn\{\mu_j:j\in J_0\}\le y-1.
\end{equation*}
Thus, by induction,
\begin{equation*}
|J_0|\le F(x,y-1).
\end{equation*}
On $J_1$, for each $j\in J_1$, we have $\mu_j(t)>0$ and
\begin{equation*}
0=(C_n)_{jj}=\langle \xi_j,\mu_j\rangle.
\end{equation*}
By nonnegativity, $\xi_j(t)=0$. Therefore the response functions
$\{\xi_j:j\in J_1\}$ lie in the same hyperplane $H_t$. Since some response
function has nonzero $t$th coordinate, this lowers the response-function span by
one:
\begin{equation*}
\dim\spn\{\xi_j:j\in J_1\}\le x-1.
\end{equation*}
Again by induction,
\begin{equation*}
|J_1|\le F(x-1,y).
\end{equation*}
Using Eq.~\eqref{eq:F-rec},
\begin{equation*}
n=|J_0|+|J_1|
\le F(x,y-1)+F(x-1,y)
=F(x,y).
\end{equation*}
This completes the induction and proves
\begin{equation*}
m\le
F\!\left(
\dim\spn\{\xi_i\},
\dim\spn\{\mu_i\}
\right).
\end{equation*}
\end{proof}

\end{widetext}

\end{document}